\documentclass{aa}  
\usepackage{graphicx,natbib,psfig}  
\newcommand{\ltsima} {$\; \buildrel < \over \sim \;$}  
\newcommand{\gtsima} {$\; \buildrel > \over \sim \;$}  
\newcommand{\lta} {\lower.5ex\hbox{\ltsima}}  
\newcommand{\gta} {\lower.5ex\hbox{\gtsima}}

\defcitealias{capetti05}{Paper I}
\defcitealias{paper2}{Paper II}
\begin{document}  

\title{The host galaxy/AGN connection in nearby early-type galaxies
\thanks
{Based  on observations obtained at
the  Space  Telescope Science  Institute,  which  is  operated by  the
Association of  Universities for Research  in Astronomy, Incorporated,
under NASA contract NAS 5-26555.}.}
\subtitle{A new view of the origin of the radio-quiet/radio-loud dichotomy?}
  
\titlerunning{A new view of the origin of the radio-quiet/radio-loud dichotomy?}
\authorrunning{A. Capetti and B. Balmaverde}
  
\author{Alessandro Capetti
\inst{1}
\and    
Barbara Balmaverde \inst{2}}    
   
\offprints{A. Capetti}  
     
\institute{INAF - Osservatorio Astronomico di Torino, Strada
  Osservatorio 20, I-10025 Pino Torinese, Italy\\
\email{capetti@to.astro.it}
\and 
Universit\'a di Torino, Via Giuria 1, I-10125, Torino, Italy\\
\email{balmaverde@ph.unito.it}}

\date{}  
   
\abstract{This is the third in a series of three papers exploring the
connection between  the multiwavelength 
properties of AGN in nearby early-type galaxies
and the characteristics of their hosts. 
Starting from an initial sample of 332 galaxies, 
we selected 116 AGN candidates requiring the detection 
of a radio source with a flux limit of $\sim$ 1 mJy,
as measured from 5 GHz VLA observations.
In \citetalias{capetti05}
we classified the objects with available archival HST images into
``core'' and ``power-law'' galaxies, 
discriminating on the  basis of the nuclear slope of their brightness
profiles. We used HST and Chandra data to isolate the nuclear
emission of these galaxies in the optical and X-ray bands, 
thus enabling us (once combined with the radio data) to study the
multiwavelength behaviour of their nuclei.
The properties of the nuclei hosted by the
29 core galaxies were presented in \citetalias{paper2}. 
Core galaxies invariably host a radio-loud nucleus,
with a median radio-loudness of Log R = 3.6 and 
an X-ray based radio-loudness parameter of Log R$_{\rm X}$ = -1.3.

Here we discuss the properties of the nuclei of the 22 ``power-law'' galaxies.
They show a substantial excess of optical and X-ray emission
with respect to core galaxies at the same level of radio luminosity.
Conversely, their radio-loudness parameters, Log R $\sim$ 1.6 and
Log R$_{\rm X} \sim$ -3.3, are similar to those measured
in Seyfert galaxies. Thus the radio-loudness 
of AGN hosted by early-type galaxies appears to be univocally 
related to the host's brightness profile: radio-loud AGN are
only hosted by core galaxies, while radio-quiet AGN are found only
in power-law galaxies. 

The brightness  profile  is determined  by  the galaxy's  evolution,
through its  merger history; our results suggest that  the same
process sets the AGN flavour. In this scenario,
the black holes hosted by the merging galaxies 
rapidly sink toward the centre of the newly formed object, 
setting its nuclear configuration, described  by e.g.
the total  mass, spin, mass  ratio, or separation  of the SMBHs.
These parameters
are most likely at the origin of the different levels of
the AGN radio-loudness.

This connection might open 
a new path toward understanding  the origin of the
radio-loud/radio-quiet  AGN dichotomy and 
provide us with a further tool
for exploring  the co-evolution of galaxies  and supermassive black
holes.

} \maketitle

\section{Introduction.}
\label{intro}
Despite the fundamental breakthroughs in 
our understanding of the nuclear regions of
nearby galaxies, such as the ubiquitous presence of supermassive
black holes \citep[e.g.][]{kormendy95}, we still lack a clear
picture of the relationship between AGN and host galaxies.
For example, while spiral galaxies preferentially harbour radio-quiet
AGN, early-type galaxies host both radio-loud and radio-quiet
AGN. Similarly, radio-loud AGN are generally associated with the most
massive SMBH as there is a median shift between the radio-quiet and
radio-loud distribution, but both distributions are broad and overlap
considerably \citep[e.g.][]{dunlop03}.
Nonetheless, recent developments 
provide us with a  new framework in  which to explore
the classical issue of the connection between host galaxies and AGN.

With this aim, we focused on two samples of nearby early-type galaxies
for which extensive multiwavelength data from VLA, HST, and
Chandra observations are available. This enabled us to
study the multi-band behaviour of their nuclei in detail 
and, at the same time, to extract the structural parameters 
describing the host galaxies.
More specifically we analysed the samples of early-type galaxies 
studied by \citet{wrobel91b}  and \citet{sadler89}, both observed with
the VLA at  5 GHz with a  flux limit  of
$\sim$  1 mJy. \citet{wrobel91a} extracted  a northern
sample  of  galaxies from  the  CfA  redshift survey  \citep{huchra83}
satisfying  the following  criteria for a total
of  216 galaxies: (1)  $\delta_{1950} \geq  0$, (2)
photometric magnitude  B $\leq$ 14; (3) heliocentric  velocity $\leq$ 3000
km s$^{-1}$; and (4) morphological  Hubble type T$\leq$-1.  
\citet{sadler89} selected  a similar southern sample
of 116 E and S0 with  $-45 \leq \delta \leq -32$. 
The only difference between
the  two  samples  is  that  \citeauthor{sadler89} did  not  impose  a
distance limit.   Nonetheless,  the  threshold in  optical  magnitude
effectively limits the  sample to a recession velocity  of $\sim$ 6000
km s$^{-1}$.

In \citet[ hereafter Paper I]{capetti05}, we selected the 116 
galaxies detected in these VLA surveys to boost the fraction
of AGN with respect to a purely optically selected sample.
We used archival HST observations, available for 65 objects, to study their 
surface brightness profiles and to separate these early-type 
galaxies following the Nukers scheme \citep{faber97} 
rather than the traditional morphological classification (i.e. into E and
S0 galaxies). Galaxies are then 
separated on the basis of the logarithmic 
slope $\gamma$ of their nuclear brightness profiles in the two classes
of ``core'' ($\gamma \leq 0.3$) and ``power-law'' ($\gamma \geq 0.5$)
galaxies. For 51 galaxies 
the surface brightness profile is sufficiently well-behaved 
to yield a successful classification.

In the second paper, \citet[ hereafter Paper II]{paper2},
we focused on the 29 ``core'' galaxies (hereafter CoreG).
We found that core galaxies invariably host a radio-loud nucleus.
Their radio-loudness parameters $R$ is on average 
Log R = Log (L$_{5\rm {GHz}}$ / L$_{\rm B})$$\sim$ 3.6.
The X-ray data provide a completely independent
view of their multiwavelength behaviour, leading to the same result,
i.e. a large X-ray deficit, at the same radio luminosity, 
when compared to radio-quiet nuclei.
Considering the multiwavelength nuclear diagnostic planes, 
the optical and X-ray nuclear luminosities are 
correlated with the radio-core power reminiscent of the behaviour
of low luminosity radio-galaxies (LLRG). The inclusion of core galaxies
indeed smoothly extends the correlations reported for LLRG 
\citep{balmaverde06} 
by a factor of $\sim 1000$  toward lower luminosities, down to 
a radio luminosity of $\nu$ L$_r \sim 10^{36}$ erg s$^{-1}$, 
covering a combined range of 6
orders of magnitude.
A strong similarity between core-galaxies and classical low-luminosity
radio-galaxies emerges, including the ``core'' classification, the
presence of collimated radio-outflows, and 
the distribution of black-hole masses.
Core galaxies and LLRG thus appear to be drawn
from the same population of early-type galaxies;
core-galaxies represent from the point of view of
the level of nuclear activity a miniature version of the radio-galaxies.

Here we discuss the properties of 
the sub-sample formed by the 19 power-law galaxies
(hereafter PlawG) and of the 3 galaxies with an intermediate
nuclear slope, i.e. with $0.3 < \gamma < 0.5$.

We have adopted a Hubble constant H$_{\rm o}=75$ km  s$^{-1}$ Mpc$^{-1}$.

\begin{table*}

\caption{Summary of the available Chandra and HST data.}

\label{tab1}

\begin{tabular}{l | c c c | l r }

\hline		      	    
\hline		      	    
Name &\multicolumn{3}{|c|} {Chandra data summary} & \multicolumn{2}{c}{HST data summary}\\
           &  Obs. Id & Exp. time [ks] & Ref. & Image & $\nu$ F$_{0}$   \\
\hline		      	    
UGC5617   &  860     & 46.9	  & (2)	  &  WFPC2/F702W     & $<$5.46E-13  \\
UGC5663   &  2926    & 9.7	  & (3)	  &  WFPC2/F702W     &    5.80E-13  \\
UGC5959   &  --      & --	  & --	  &  WFPC2/F814W     &    5.74E-13  \\
UGC6153   &  2482    & 89.4	  & (4)	  &  NICMOS/F160W    &    2.73E-11  \\
UGC6860   &  --      & --	  & --	  &  WFPC2/F814W     & $<$2.81E-13  \\
UGC6946   &  --      & --	  & --	  &  WFPC2/F547M     &    3.10E-12  \\
UGC6985   &  --      & --	  & --	  &  WFPC2/F814W     & $<$1.21E-12  \\
UGC7005   &  --      & --	  & --	  &  WFPC2/F547M     & $<$1.64E-13  \\
UGC7103   &  1587    & 15.0	  & (1)	  &  NICMOS/F160W    &    1.27E-11  \\
UGC7142   &  1617    & 2.8	  & (5)	  &  WFPC2/F606W     &    5.37E-13  \\
UGC7256   &  397     & 1.8	  & (6)	  &  WFPC2/F814W     &    1.06E-12  \\
UGC7311   &  --      & --	  & --	  &  WFPC2/F702W     &           -- \\
UGC7575   &  2883    & 25.4	  & (7)	  &  WFPC2/F814W     &           -- \\
UGC7614   &  2927    & 9.9	  & (1)	  &  WFPC2/F814W     & $<$7.61E-13  \\
UGC8355   &  --      & --	  & --	  &  WFPC2/F702W     &    6.90E-13  \\
UGC8499   &  --      & --	  & --	  &  WFPC2/F702W     & $<$1.00E-13  \\  
UGC8675   &  415     & 1.8	  & (1)	  &  NICMOS/F160W    &    2.26E-12  \\
UGC9692   &  --      & --	  & --	  &  WFPC2/F814W     & $<$5.26E-13  \\
UGC10656  &  --      & --	  & --	  &  WFPC2/F702W     & $<$3.16E-13  \\
UGC12759  &  --      & --	  & --	  &  NICMOS/F160W    &    1.81E-12  \\
\hline	      	        	  
NGC1380   &  --      & --	  & --	  &  NICMOS/F160W    & $<$6.23E-13 \\  
NGC6958   &  --      & --	  & --	  &  WFPC2/F814W     &    1.90E-12 \\
\hline						     
\end{tabular}

(1) this work, (2) \citet{george01}, (3) \citet{filho04},  
(4)\citet{netzer02}, (5) \citet{terashima03}, (6) \citet{ho01}, (7) \citet{machacek04}. 
\end{table*}

\section{Basic data and nuclear luminosities}
\label{nuc}

The basic
data for  the power-law galaxies, namely the recession velocity  
(corrected for Local Group infall onto Virgo), the K
band  magnitude from  the Two  Micron  All Sky  Survey (2MASS),
the galactic extinction, and the total and
core radio fluxes were given in \citetalias{capetti05}.
In the following three subsections, 
we derive and discuss the measurements for the
nuclear sources in the optical, X-ray, and radio bands.

\subsection{Optical nuclei.}

\begin{figure*}
\centerline{
\psfig{figure=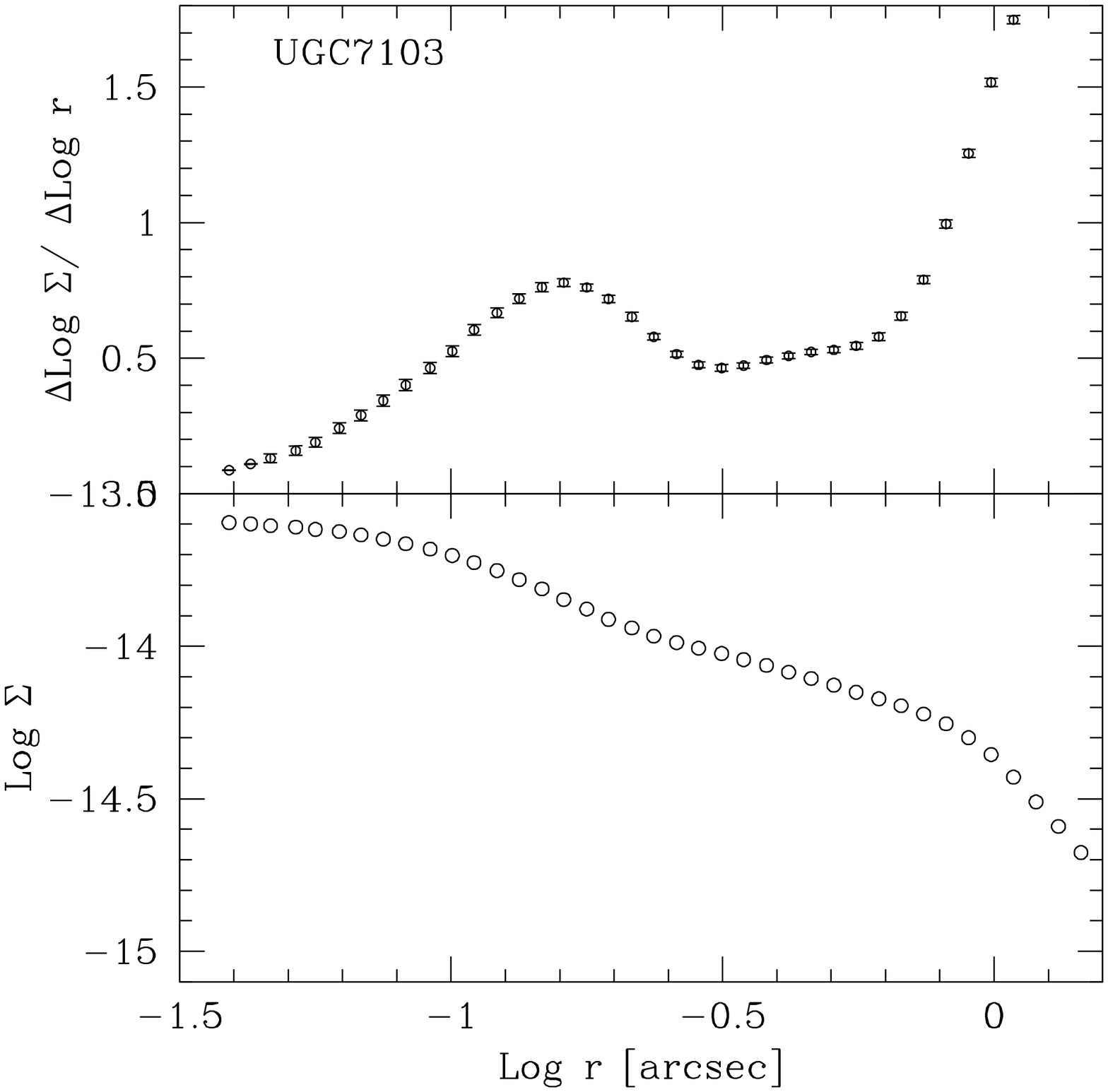,width=0.33\linewidth}
\psfig{figure=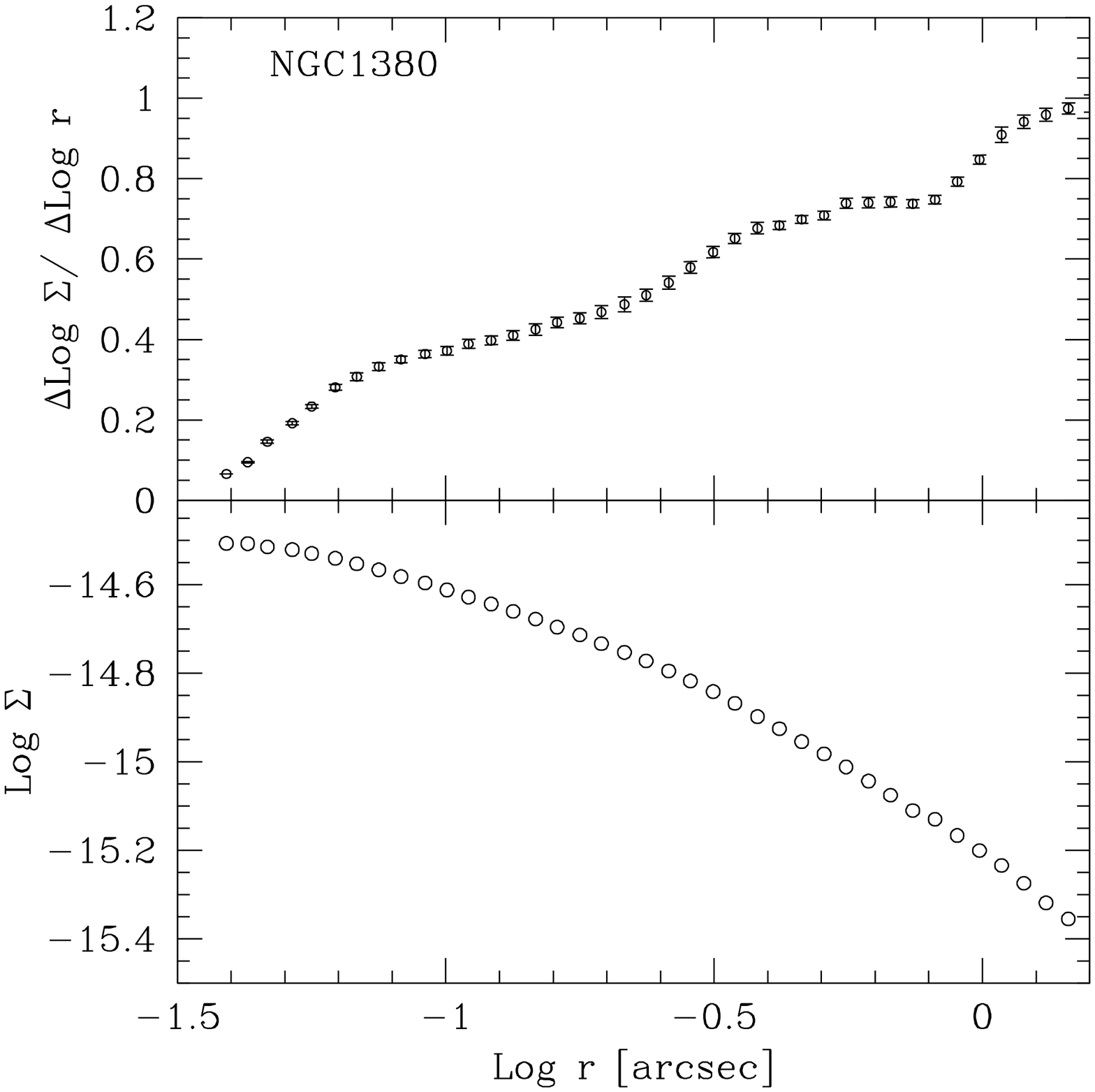,width=0.33\linewidth}
\psfig{figure=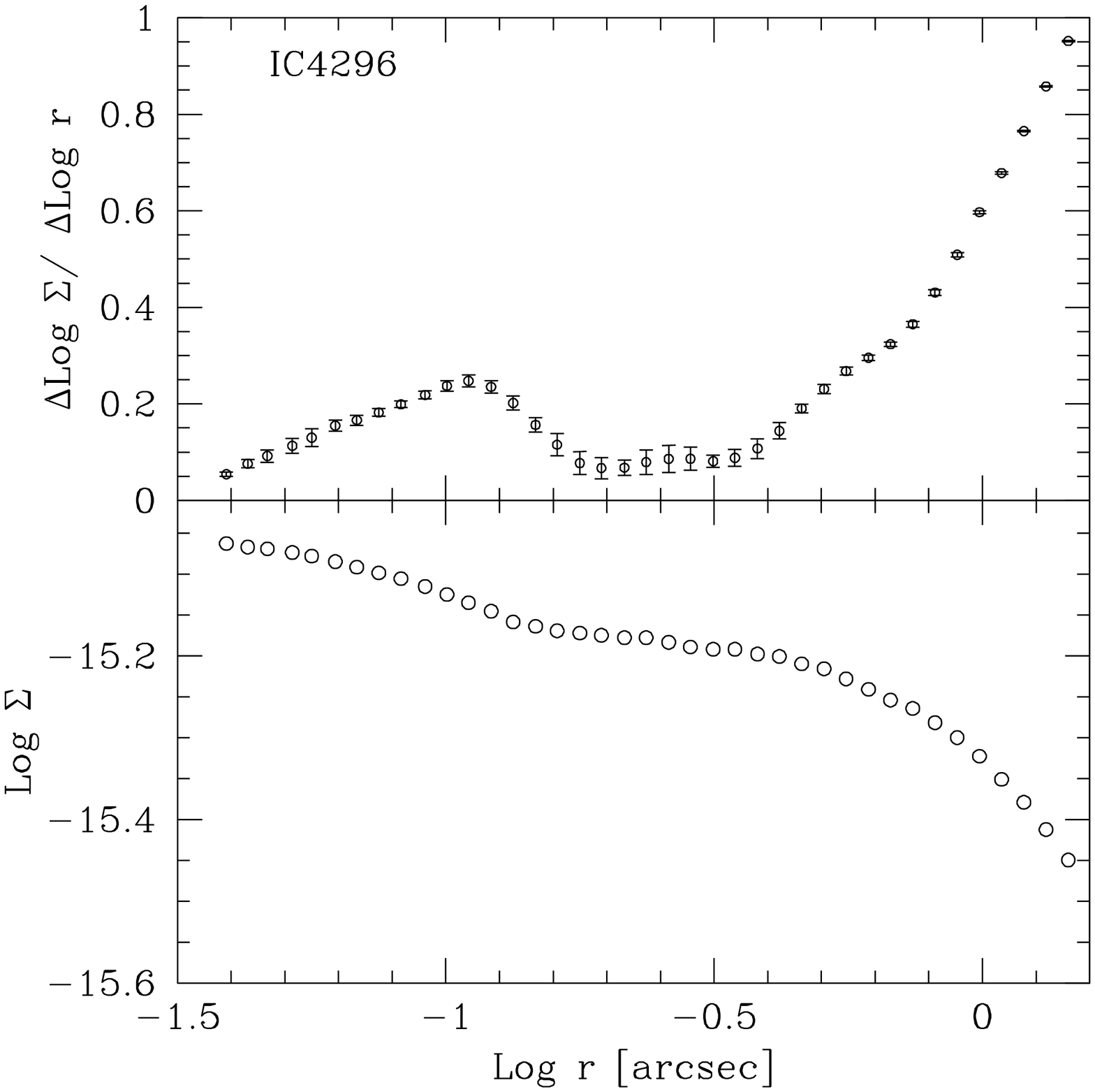,width=0.33\linewidth}
}
\caption{\label{hstnuc} Brightness profile and its derivative for 
two PlawG: UGC~7103 (left panel) clearly shows the presence of a 
nuclear source, with 
a highly significant increase in the derivative at about $r=0\farcs3$.
In NGC~1380 (middle panel) no up-turn in the
brightness profile is seen and it is considered as a non detection.
The profile of the nucleated CoreG IC~4296 is shown as a comparison in the
right panel. All 3 profiles presented are derived 
from the same filter/instrument combination (NICMOS/F160W) HST images.
}
\end{figure*}

Following the strategy outlined in \citetalias{paper2} 
for establishing whether an optical nuclear source is present at 
the centre of the objects of our sample, we relied on 
analysing the derivative of the surface brightness profile.
In particular we looked for the characteristic up-turn 
that unresolved nuclear sources 
cause in the nuclear brightness profile, as opposed to
the smooth decrease of the slope toward the centre
when only a diffuse galactic component is present.
Thus, we  evaluated the  derivative  of  the
brightness profile in a log-log representation,  
combining the brightness 
measured over two adjacent points on each side of the radius of interest,
yielding a second-order accuracy. We then looked for  
the presence of a nuclear up-turn in the derivative 
requiring a difference larger than
3 $\sigma$ from the slopes at the local minimum and maximum
for a nuclear detection.

This a rather conservative definition, since 
i) the region over which the up-turn is detected 
extends over several pixels, while we only consider the
peak-to-peak difference, and ii) 
a nuclear source can still be
present but its intensity might not be sufficient to compensate for the
downward trend of the derivative sets by the host galaxy, leading only
to a plateau in the derivative. 

In Fig. \ref{hstnuc} we  report three examples. 
The brightness profile of UGC~7103 (left panel)
clearly shows the presence of a nuclear source, with 
a highly significant increase in the derivative at about $r=0\farcs3$.
In NGC~1380 (middle panel) no up-turn in the
brightness profile is seen, it is so considered as a non detection.
The profile of the nucleated CoreG IC~4296 is shown as a comparison in the
right panel.

In five cases, the  optical images are strongly  affected by
the  presence of  dust absorption  features. They  not only  prevent 
detailed study of their brightness  profiles, but they also plague the
measurements of the nuclear optical sources. For these objects we used
the infrared HST  images at 1.6 $\mu$m to  measure their nuclei. 
This approach appears to be reasonably justified within  the context of a
multiwavelength study spanning more than seven orders of magnitude in
frequency, where  the difference in  frequency between optical  and IR
data is only marginal.
We also note that the fraction of nuclei measured from IR data
is rather small in both  
sub-samples of early type galaxies ($\sim 24$\%). 

Nonetheless, the use of IR measurements as a
substitute for the optical nuclear fluxes is
a potential cause of concern\footnote{In Paper I we discussed a 
similar issue, showing 
that the Nuker classification is independent of the observing band.}.  
First of all, this is related to the lower resolution of the
NICMOS images, since at 1.6 $\mu$m the diffraction limit of HST is
$\sim 0\farcs17$, a factor $\sim$2 larger than in the optical images.
Consequentely,
the detection threshold of an IR nucleus is higher than in the optical
band. However, only one PlawG (and only one CoreG in Paper II) 
is undetected in the IR images, indicating that this is not a
significant issue. Similarly, the contrast against the galaxy's
background might vary from the optical to the IR.  
However, the  ratio between the optical (e.g. I band)  and H  band for  the host
galaxies, in  terms of  $\nu $F$_\nu$, is $\sim 0.9$ for early type
galaxies \citep{mannucci01}. Even adopting a wide
range of values for the nuclear spectral index, 
e.g. $\alpha \sim$ 0 -- 2, the contrast between AGN
and galaxy  in the optical and IR  images would differ by  less than a
factor of 2 (in either direction) in the two bands.

We also  investigated the potentially worrying issue  of how the
fluxes of  the 5 PlawG nuclei measured  in the IR appear to  be on average
higher that  those obtained from the  optical images. In  2 cases, the
optical nucleus  has a  luminosity (although measured  with relatively
large uncertainties due  to the presence of dust) within  30 \% of the
IR nucleus.   Three of  them instead  have X-ray data, and  their X-ray
nuclei are among the brightest in the sample, similar to what is seen
for the IR data.  This indicates that their nuclei are indeed brighter
than average, regardless  of the observing band used  to measure their
fluxes, and that this is not a bias introduced by the use
of the NICMOS images. This effect 
is  probably due to the connection  between the presence
of  dust structures  and nuclear  activity  \citep[e.g.][]{lauer05},
since 
we used IR  data for the  dustier galaxies that  are more likely  to host
bright AGN.
  
Adopting the strategy described above, we identified a
nuclear source in the HST images
in 11 out of 22 objects, four of them from IR data. 
We measured the nuclear fluxes (reported in 
Table \ref{tab1}) as in \citetalias{paper2}, i.e.
integrating in a circular aperture with
radius set at the location of the up-turn and as background region 
a circumnuclear annulus, 0.1$\arcsec$ in width. 
In 9 objects we did not find a nuclear upturn in the brightness profile,
so they are considered as non detections. 
For the undetected nuclei we set as upper limits the light excess
with respect to the starlight background
within a circular aperture 0.1\arcsec\ in diameter.
In the remaining 2 objects (UGC~7311 and UGC~7575) 
the central regions are hidden by a kpc scale dust
structure (see Fig. 2 in \citetalias{capetti05}) 
so no measurement can be performed.

We derived all the
luminosities referring to 8140 \AA\ (see Table \ref{lum}) 
and adopting a spectral 
index\footnote{We define the spectral index $\alpha$ with the spectrum
in the  form F$_\nu  \propto \nu^{-\alpha}$} $\alpha_o = 1$.

\subsection{X-ray nuclei.}

For the measurements of the X-ray nuclei we concentrated 
on the Chandra measurements, as this telescope provides 
the best  combination of
sensibility and resolution necessary to detect the faint nuclei expected in
these weakly active galaxies.
Data for 9 power-law galaxies 
are available in the Chandra public archive.
In Table \ref{tab1} we give a summary of the available
Chandra data, while references and details for the X-ray 
observations and analysis are presented in Appendix \ref{notes}. 

When possible, we used the results of the analysis of the X-ray
data from the literature. These are available for 6 sources,
and in all cases they show the presence of a nuclear source, whose
luminosity we rescaled to our adopted distance 
and converted to the 2-10 keV  band, using  the
published power law index. 
We also considered the Chandra archival data for the 3 unpublished
objects.
We analysed these
observations using the same strategy as in
\citet{paper2}. 
For 2 objects (namely UGC~4111 and UGC~8675) we obtained  a detection of a
nuclear power-law source.
The data for one object (UGC~7614) do not lead to the detection
of a X-ray nucleus. Details of the results are given in Appendix \ref{notes},
while the X-ray luminosities for all objects are given in Table \ref{lum}.

\subsection{Radio nuclei.}

The radio data of our sample 
are drawn from the surveys by \citet{wrobel91b} 
and \citet{sadler89}. Since they were obtained
at a resolution of $\sim$ 5\arcsec,
they do not usually have a resolution sufficient, in the case of PlawG,
to separate the core
emission from any extended structure. 
Therefore, it is possible 
that these VLA data might overestimate the core flux.

We thus searched the literature  for radio-core measurements  obtained at
a higher  resolution  (and/or  higher  frequency) than the VLA-survey 
data so as to improve  the  estimate  of the  core  flux  density.
Better measurements from  VLBI data  or from  higher  frequency/resolution VLA
data are available for most PlawG  (17 out of 22) 
\footnote{More specifically the radio-core
fluxes are taken from \citet{nagar00} (15 GHz VLA data at 0\farcs2 resolution),
\citet{nagar99} (8.4 GHz VLA data at 0\farcs3), 
\citet{nagar02} (15 GHz VLA data and 5 GHz VLBI data), 
\citet{krajnovic02} (8.4 GHz at the VLA),
\citet{filho02} (VLBI data at 5 GHz),
\citet{filho04} (VLA data at 8.4 and 5 GHz), and
\citet{slee94} (PTI 8.4 GHz data).}.

\begin{figure}
\centerline{
\psfig{figure=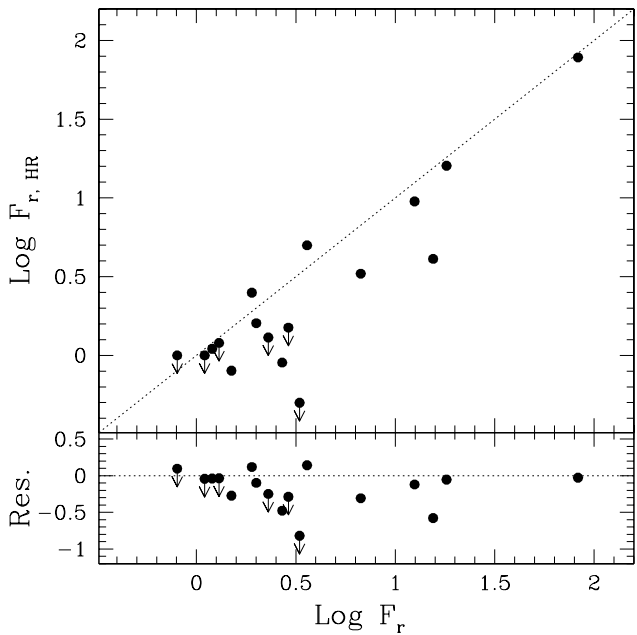,width=1.00\linewidth}
}
\caption{Radio-core flux density for PlawG obtained at 
5 GHz with the VLA (used in this work) compared to higher resolution data.
See text for the references.
The dotted line corresponds to equality.}
\label{cfr}
\end{figure}

These observations detected in 11 out of the 17 objects with available data. 
In Fig. \ref{cfr} we
compare the radio  core flux density used in our  analysis 
with the observations made at higher resolution.
Overall there  is substantial agreement between the two datasets.
Considering only the detections,
the median ratio between the high and low resolution data is $\sim$0.80, with
only one object whose offset is as large as a factor 3. 
Five of the upper limits are located close to the line of equal
fluxes, while one of the undetected objects (UGC~5663) has
a deficit larger than a factor of 6.
Since  these data are  highly inhomogeneous and incomplete,
and given the general  agreement with the  5 GHz VLA measurements, we
prefer to retain Wrobel's and  Sadler's values.
Nonetheless, in the rest of our analysis we 
always checked whether our results would be significantly affected
using these higher resolution core fluxes.

\section{The multiwavelength properties of the nuclei of power-law galaxies.}
\label{nuclei}

\begin{table*}
\caption{Power-law galaxies data.}

\label{lum}

\centering
\begin{tabular}{l c c c c c c}
\hline \hline
Name     & Log L$_{X}$ &   Log $\nu L_{o}$  &  Log $\nu L_{core}$ &   Log L$_{[O III]}$  &  M$_K$ & Log $(M_{BH}/M_{\odot})$  \\
\hline  
UGC5617   & 40.3    & $<$ 40.51 & 36.88 &     38.75   &  -22.80  &  8.07 \\
UGC5663   & 38.8    &  40.57    & 36.87 &     38.59   &  -23.58  &  8.32$^{b}$ \\
UGC5959   & --      &  40.54    & 37.16 &     38.95   &  -23.73  &  8.43 \\
UGC6153   & 42.4    &  42.11    & 38.14 &     40.79   &  -24.43  &  7.42 \\
UGC6860   & --      &  $<$40.15 & 36.38 &     38.28   &  -23.98  &  7.89 \\
UGC6946   & --      &  41.40    & 38.16 &     39.42   &  -23.80  &  8.87 \\
UGC6985   & --      &  $<$40.56 & 36.31 &      --     &  -23.39  &  7.93 \\
UGC7005   & --      &  $<$40.37 & 36.94 &     39.08   &  -24.20  &  8.03 \\
UGC7103   & 39.4    &  40.85    & 36.39 &     38.45   &  -23.08  &  7.60 \\
UGC7142   & 39.9    &  40.44    & 36.96 &     38.71   &  -23.05  &  8.25 \\
UGC7256   & 40.2    &  40.57    & 37.33 &     39.01   &  -23.73  &  7.76 \\
UGC7311   & --      &  --       & 37.07 &      --     &  -23.76  &  8.30 \\
UGC7575   & 39.3    &  --       & 36.00 &     37.71   &  -23.07  &  7.71 \\
UGC7614   & $<$40.4 &  $<$40.47 & 36.15 &  $<$37.85   &  -24.03  &  7.84$^{b}$  \\
UGC8355   & --      &  41.27    & 36.69 &      --     &  -22.75  &   --  \\
UGC8499   & --      &  $<$40.37 & 37.09 &      --     &  -23.95  &  8.09 \\
UGC8675   & 40.6    &  40.33    & 36.34 &     39.66   &  -22.56  &   6.19 \\
UGC9692   & --      &  $<$40.41 & 36.65 &     37.78   &  -23.85  &  8.63 \\
UGC10656  & --      &  $<$40.96 & 37.05 &      --     &  -24.00  &  7.63 \\
UGC12759  & --      &  40.49    & 36.93 &     39.51   &  -23.39  &  6.61 \\
\hline	               	       	                  
NGC1380   &  --     &  $<$39.96 & 36.68 &      --     &  -24.80	 &  8.30 \\
NGC6958   &  --     &  41.46    & 38.10 &       --    &  -24.27  &  8.00 \\
\hline		   										 
\end{tabular}	   										

Columns description: (1) UGC name, (2) nuclear X-ray luminosity (2-10 keV)[erg s$^{-1}$],
 (3) nuclear optical luminosity (8140 \AA) [erg s$^{-1}$] corrected for absorption using the galactic extinction values
in Paper I, (4) nuclear radio luminosity (5 GHz) [erg s$^{-1}$] derived from Paper I,
(5) [O III] line luminosity [erg s$^{-1}$]
from Ho et al. (1997), (6) total K band galaxy's absolute magnitude from
2MASS, (7) logarithm of black hole mass in solar unity from $^{b}$ Marconi et al. (2003) or derived using the velocity dispersion.
 
\end{table*}

\begin{figure*}
\centerline{
\psfig{figure=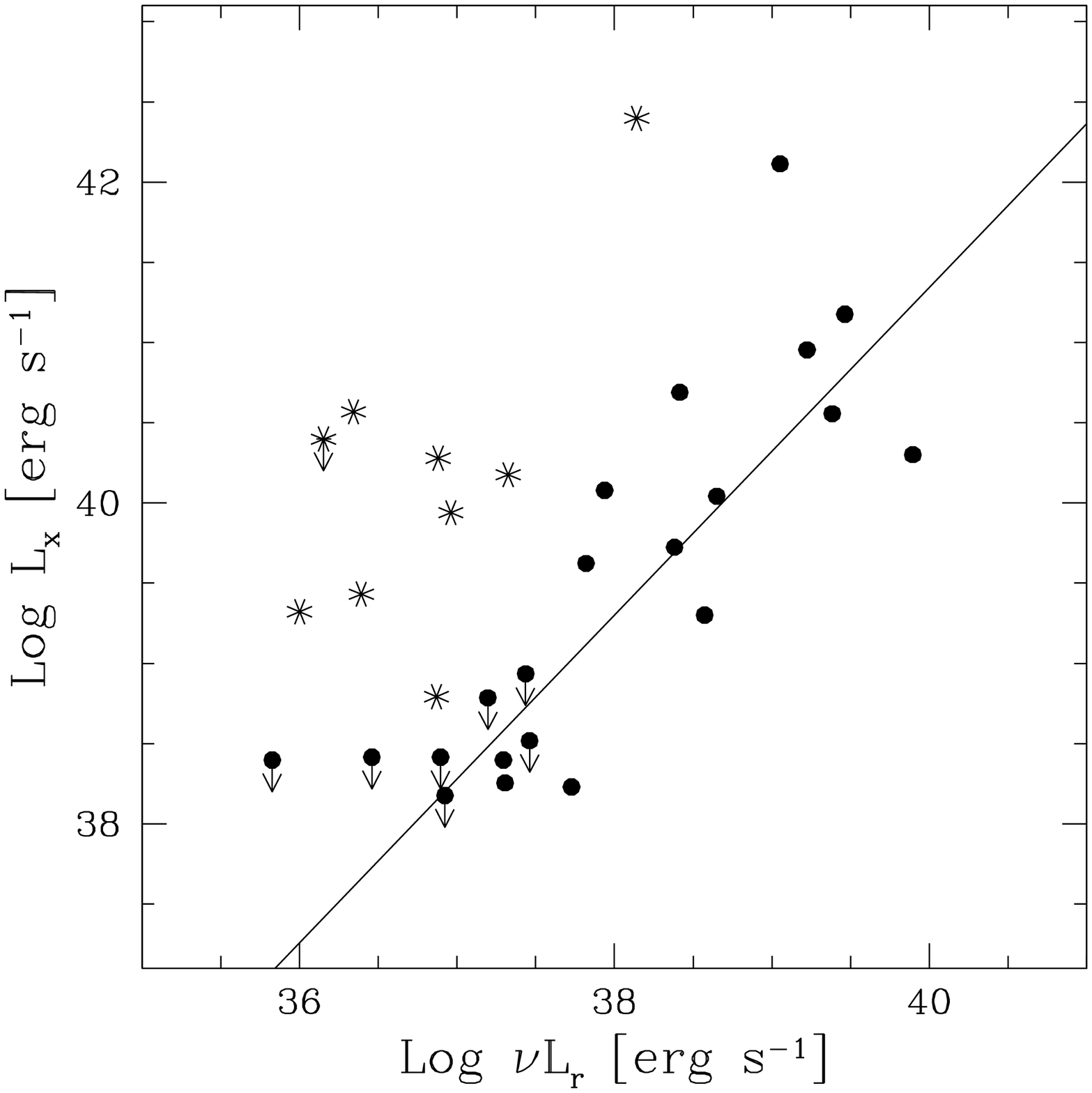,width=0.50\linewidth}
\psfig{figure=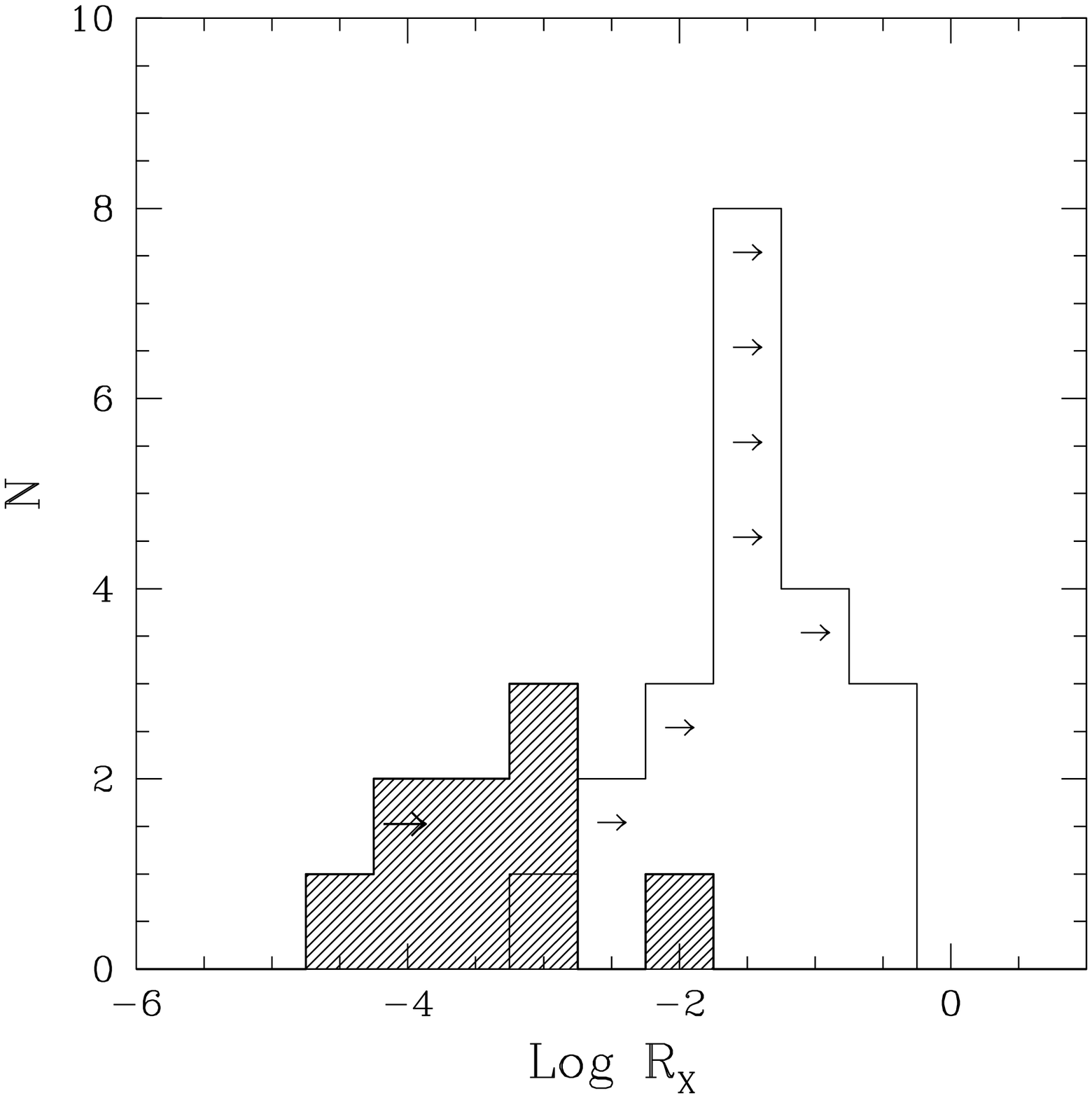,width=0.50\linewidth}}
\caption{\label{lrlx} Left panel: 
comparison of radio and  X-ray nuclear luminosity
for the sample of power-law galaxies (stars) and for the 
core galaxies (filled circles). The solid line represents the 
correlation derived in \citetalias{paper2} between the nuclear luminosities 
of the sample formed by CoreG and the 
3C/FR~I sample of low luminosity radio-galaxies.
Right panel: radio-loudness parameter
$R_X$ estimated from the ratio of radio and X-ray nuclear
luminosity, i.e. R$_{\rm X}$  = $(\nu L_{\rm r}/L_{\rm X})$,
for PlawG (filled histogram) and CoreG (empty histogram).}
\end{figure*}

Having collected the multiwavelength  information for the nuclei of our
galaxies, we  can now compare  the emission  in the  different bands. 
We detected an X-ray nuclear source in 8 out of 9 PlawG 
with available data, with only the exception of UGC~7614. 
Their representative points in the nuclear diagnostic diagram 
in which radio and X-ray luminosity are compared (see Fig. \ref{lrlx}) are 
located well above the correlation defined by the CoreG and the LLRG.
The upward offsets are 
between 0.8 and 3.3 dex, with a median of 1.8 dex.

A radio-loudness parameter based on the comparison
between radio and X-ray luminosity 
(defined as the ratio of the radio-luminosity at 5 GHz and 
the 2-10 keV X-ray luminosity, i.e. 
R$_{\rm X}$  = $\nu L_{\rm r}/L_{\rm X}$) 
has been recently introduced by \citet{terashima03}.
For PlawG we find values in the range 
Log R$_{\rm X} \sim -3 $ -- $ -4.3$
\footnote{The only exception is UGC~5663 with Log R$_{\rm X} $ = -1.9. 
However, this object is not detected in the VLBI observations 
discussed in Sect. \ref{nuc}. This indicates that 
a value of Log R$_{\rm X} $\ltsima -2.7 is probably a more accurate
estimate.}
with a median value of Log R$_{\rm X}  \sim -3.3$. 
These values are very similar to those obtained 
by \citet{terashima03} for the nuclei of Seyfert galaxies, and indeed
3 PlawG are spectroscopically classified as Seyfert by \citet{ho97}. 
Conversely the radio-to-X-ray luminosity ratios for PlawG
are substantially lower than the median value obtained for the
radio-loud nuclei of the core galaxies, for which we found
Log R$_{\rm X}  \sim -1.3$. 

Overall, the optical observations confirm this picture. 
The 11 galaxies with a detected optical nucleus are located
between 1.7 and 3 orders of magnitude above the L$_{\rm r}$ vs
L$_{\rm o}$ correlation found for CoreG
(see Fig. \ref{lrlx}, left panel).
All optical upper limits are located in the same region
of the radio vs. optical plane of the detected nuclei.
Nonetheless, the relatively large fraction of upper limits
warrant a more detailed discussion of this issue, which  will be presented
in the next sub-section.

\begin{figure*}
\centerline{
\psfig{figure=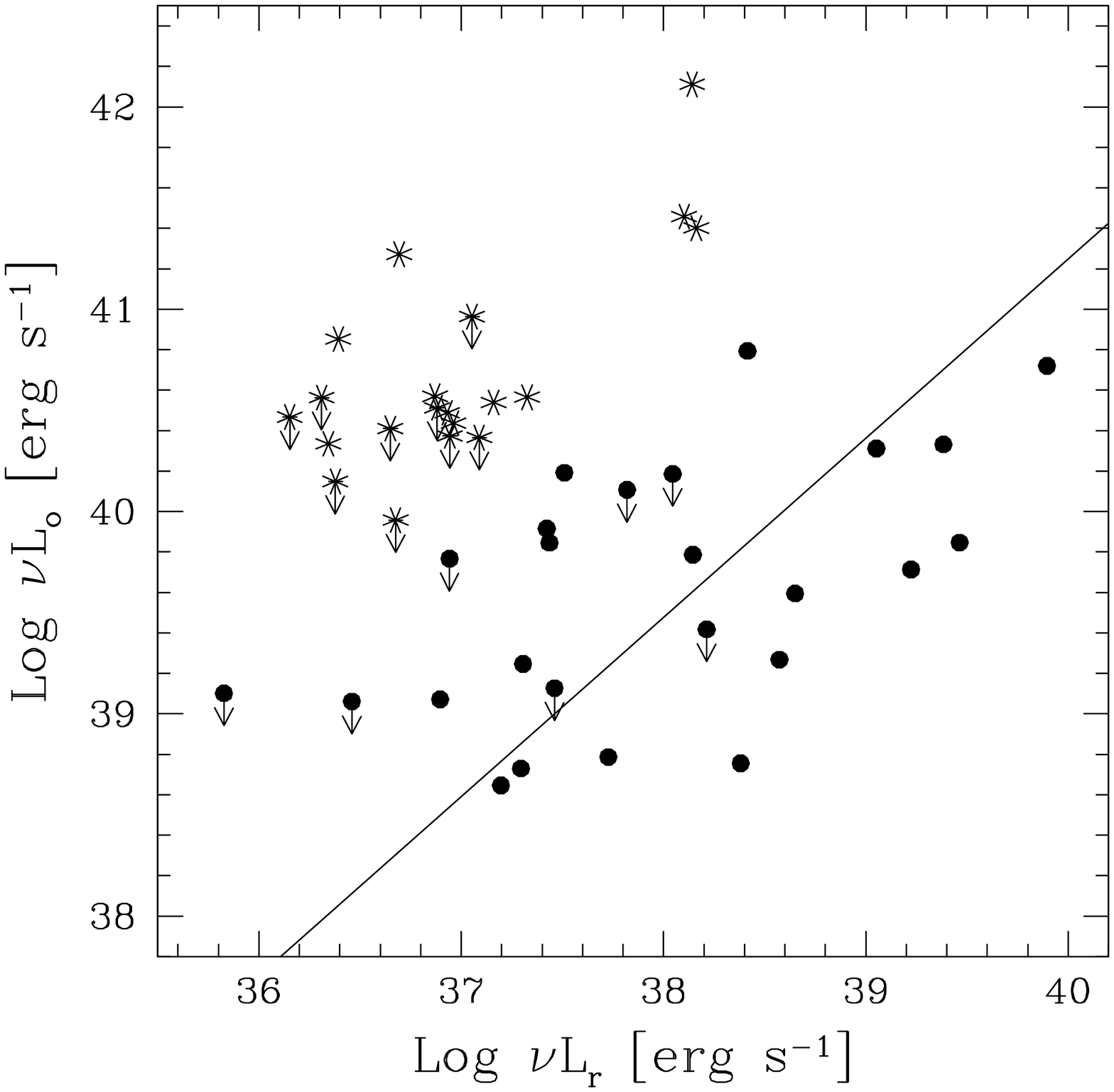,width=0.50\linewidth}
\psfig{figure=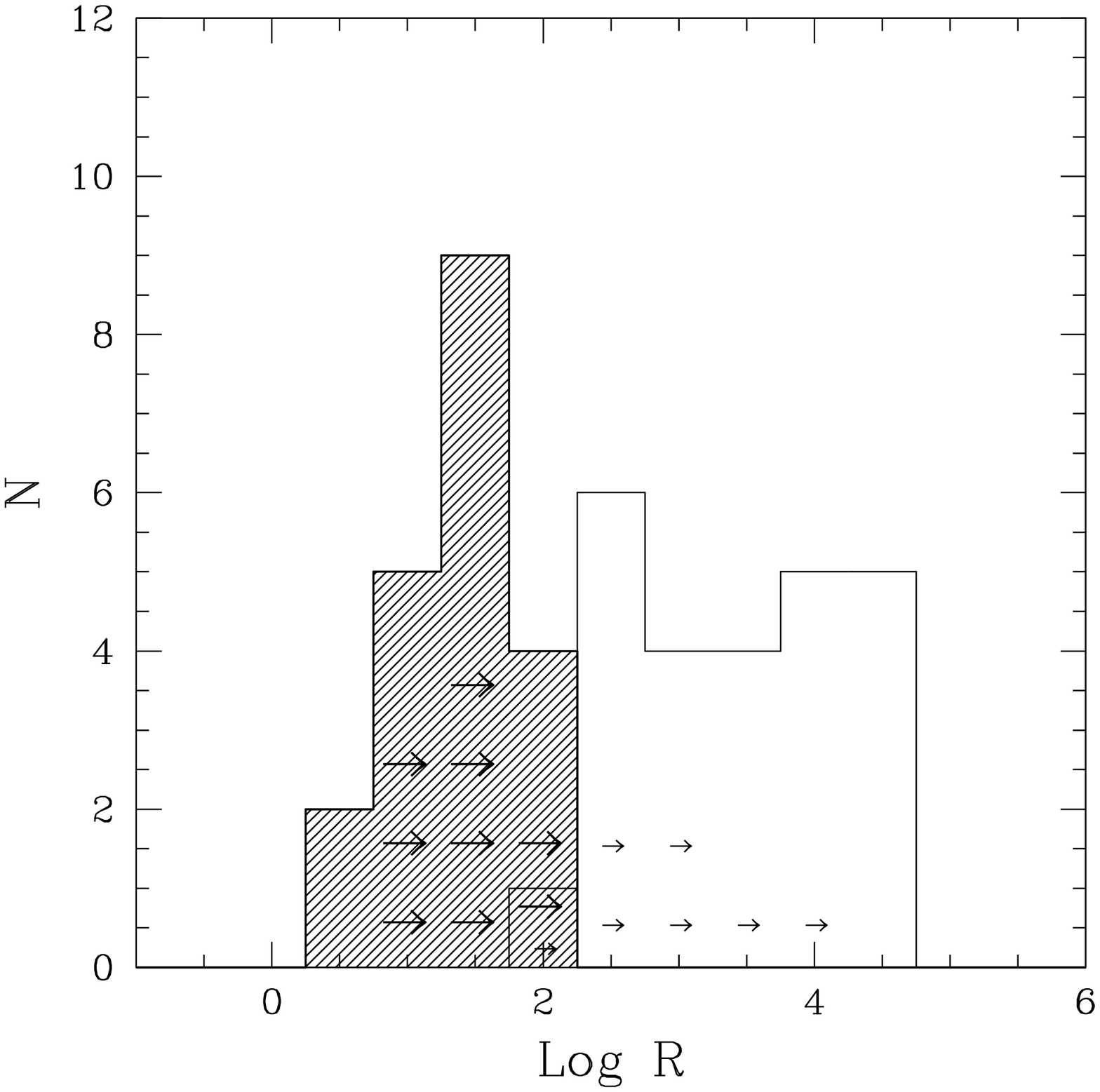,width=0.50\linewidth}
}
\caption{\label{lrlo} Same as Fig. \ref{lrlx} but using
the optical nuclear luminosity. Left panel: 
comparison of radio and optical luminosity
for the PlawG (stars) and CoreG (filled circles). 
The solid line represents the 
correlations derived in \citetalias{paper2} between the nuclear luminosities 
of the sample formed by CoreG and the 
3C/FR~I sample of low luminosity radio-galaxies.
Right panel: comparison of the radio-loudness parameter
R =  L$_{5\rm {GHz}}$ / L$_{\rm B}$ for PlawG
(filled histogram) and CoreG
(empty histogram); larger, heavier arrows mark the upper limits for
PlawG, while smaller arrows are associated with CoreG upper limits.}
\end{figure*}

In Fig. \ref{lrlo}, right panel, we compare 
the distributions of the standard radio-loudness parameter 
R =  L$_{5\rm {GHz}}$ / L$_{\rm B}$ \citep{kellermann94}
having converted the optical
luminosity to the B band with an optical spectral index of $\alpha = 1$.
It shows the presence of
an overall shift of about two orders of magnitude between the two
classes of early-type galaxies.

Here the traditional separation into
radio-loud  and radio-quiet  AGN 
\citep[Log R =  1, e.g.][]{kellermann94}
is probably inappropriate since this is based
on the observations of AGN with much higher luminosity, mostly bright
QSOs. In fact the results presented by \citet{ho01} clearly show that
there is a strong evolution of the radio-loudness
parameter with luminosity. This smoothly increases from  
a typical value of log $R \sim -1$ for
QSO with $M_B \sim -24$ to log $R \sim 0 $ -- 2  in Seyfert
nuclei, the same range covered by PlawG.
 
We conclude that both radio-loudness parameters, R and R$_{\rm X}$,
for PlawG are a factor of $\sim$ 100 smaller than in the radio-loud
nuclei of CoreG, while they are in close agreement  
with the values measured in radio-quiet AGN with a similar level of 
nuclear luminosity.

\begin{figure*}
\centerline{
\psfig{figure=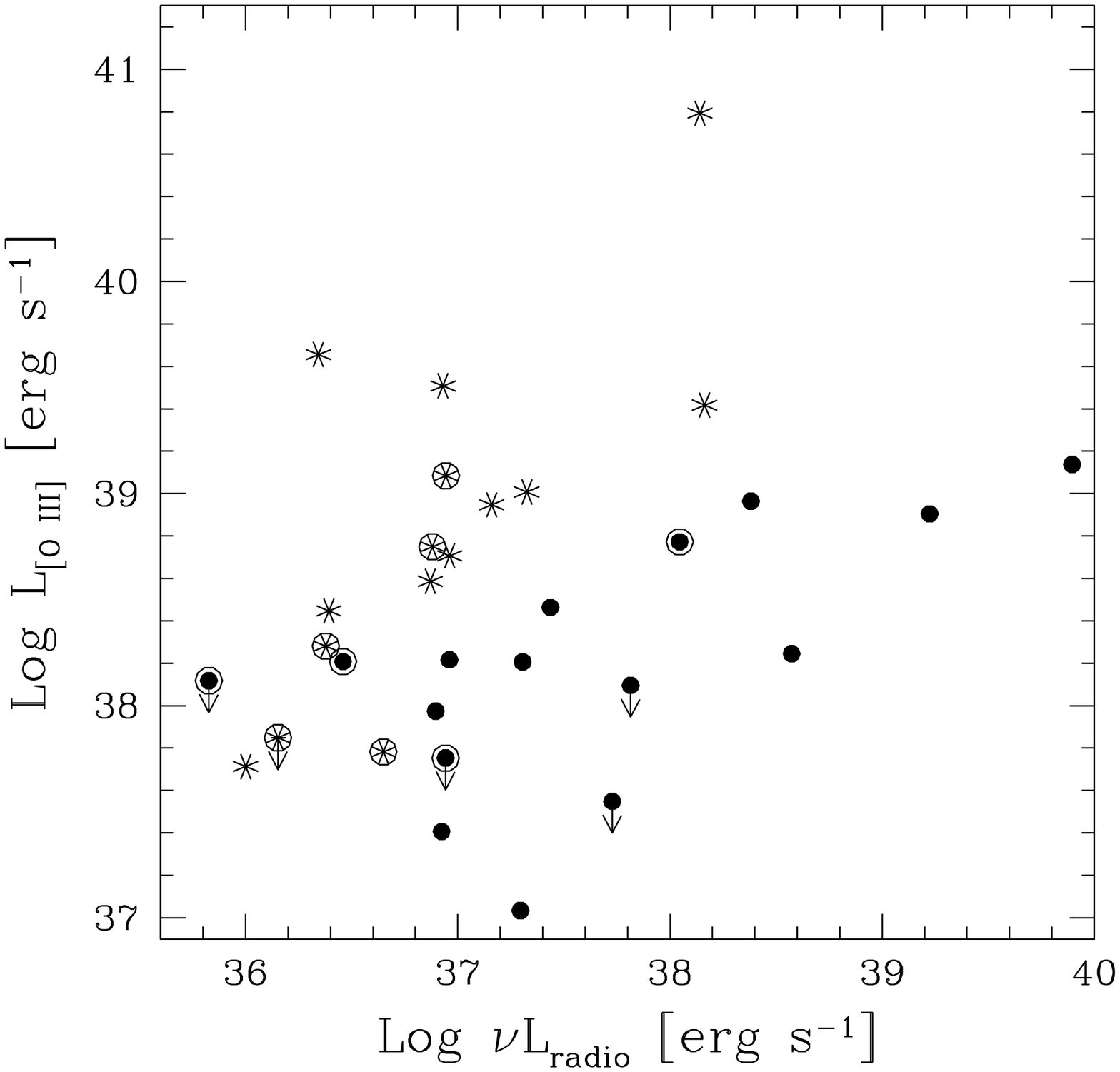,width=0.5\linewidth}
\psfig{figure=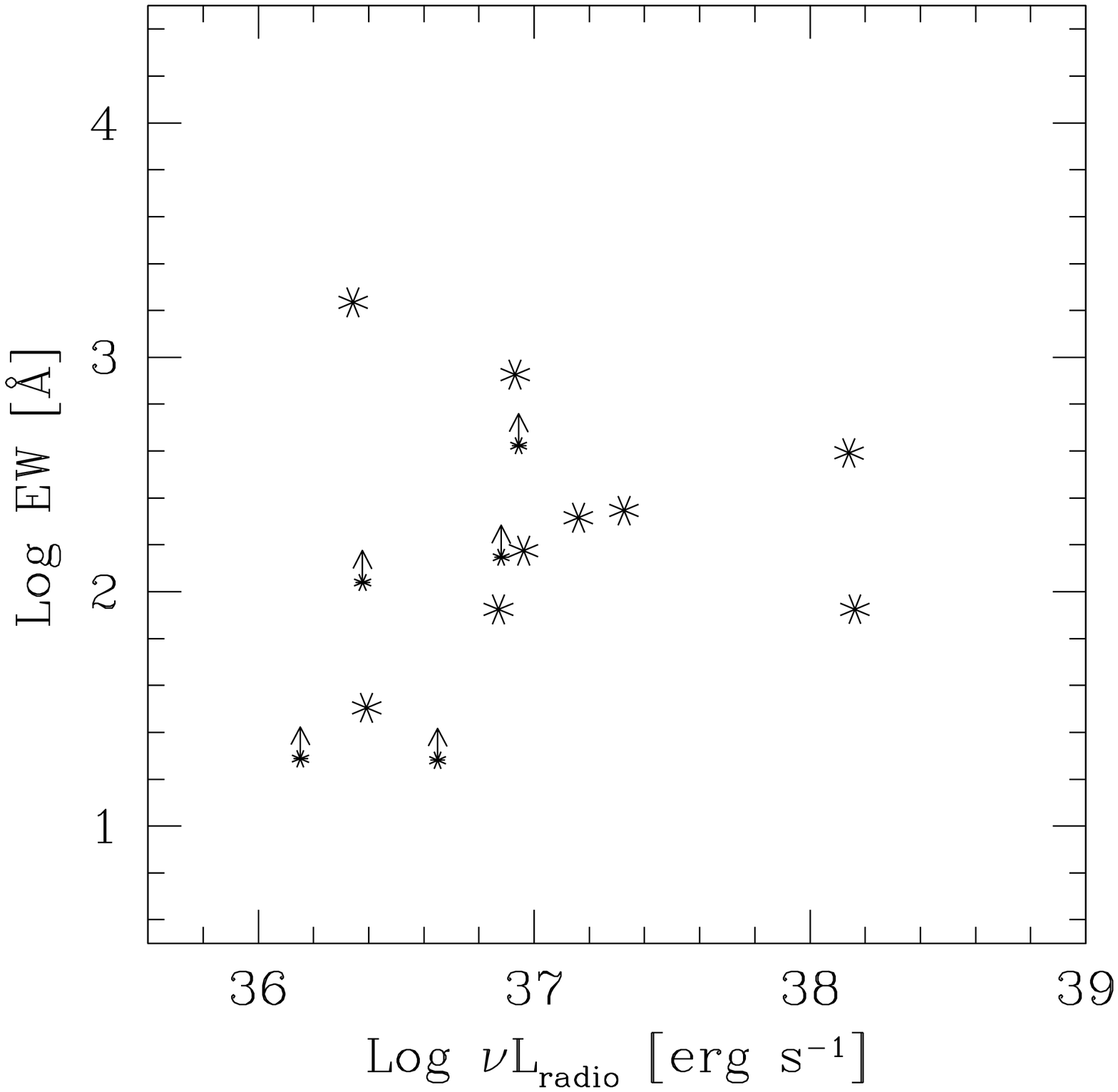,width=0.5\linewidth}
}
\caption{Left) [O III] emission line vs. radio-core luminosity 
for PlawG (stars) and CoreG galaxies (filled circles). 
Symbols enclosed in a circle mark objects with undetected optical
nuclei. Right)
Equivalent width of the [O III] emission line with respect to the
nuclear continuum luminosity.}
\label{linepl}
\end{figure*}

The comparison of the optical emission line luminosities  
provides another tool to explore the multiwavelength properties
of PlawG and CoreG nuclei. 
In the compilation of \citet{ho97} we find measurements 
of the [O III] line for most
of the objects forming the 
Northern sample (see Table \ref{lum}). In \citetalias{paper2}
we noted that CoreG extend the correlation between
radio-core  and line luminosity found for low luminosity
radio-galaxies \citep{capetti:cccriga}. Conversely, 
PlawG have a line luminosity that is highly enhanced  
with respect to CoreG at the same level of radio-luminosity
(see Fig. \ref{linepl}). This is a further indication that in PlawG
the active nucleus provides (at a given L$_r$) a substantially
larger flux of ionizing photons with respect to CoreG, 
in line with the higher luminosity 
in the optical and X-ray bands.

It is also interesting that the equivalent widths of the 
[O III] emission line (Fig. \ref{linepl}, right panel), estimated
against the HST optical nucleus, nicely fit the trend of 
EW with luminosity presented by \citep{dietrich02},
with the values measured for the PlawG again in the same range
as those of Seyfert galaxies.

We finally note that in our sample there are 3 galaxies with a cusp slope 
that is intermediate between CoreG and PlawG, $0.3<\gamma<0.5$,
namely UGC~7005, UGC~7575, and UGC~8675. Since they were not 
discussed Paper II, where only CoreG were presented, we  
included them in the present analysis. In UGC~8675, we detected
both an optical and a X-ray nucleus. The nuclear regions
of UGC~7575 are obscured by a kpc dusty disk, but this object
has a detected X-ray nucleus. Their radio-loudness
parameters are close to the average for PlawG. 
The nucleus
of UGC~7005 remains undetected in the HST images (and no X-ray data 
are available), but its radio and [O III]
luminosities (L$_{r} = 10^{36.94}$ erg s$^{-1}$ and 
L$_{\rm[O III]} = 10^{39.08}$ erg s$^{-1}$, respectively) locate this
object within the region covered by PlawG. 
We conclude that the nuclear properties of these 3 intermediate
galaxies (from the point of view of brightness profile) 
are indistinguishible from those of bona fide PlawG. 

\subsection{Undetected optical nuclei in power-law galaxies.}

As noted above, an optical nucleus is not detected
in several PlawG: in 2 cases the presence of large scale 
dust lanes prevent from exploring their optical nuclear
properties altogether, while for 9 objects we were only able
to set an upper limit. This substantial fraction is 
also related to our quite conservative definition of a nucleated
source. Despite the overall agreement between upper limits and detections,
this raises the possibility
that faint nuclei, characteristic of CoreG, might be associated
with these objects. 

In fact, the minimum detection level for a nucleus in a PlawG 
is on average substantially higher than for the CoreG,
due to the higher brightness in the nuclear regions
of PlawG\footnote{The two classes have similar brightness at the break radius,
see Fig. 10 in \citetalias{capetti05}, but the PlawG have by definition
steeper nuclear slopes, leading to higher nuclear brightness.} 
as well as to the steeper slope 
of the brightness profiles close to the centre. 
This is explained better by 
comparison of the results obtained for the analysis of the
brightness profile of a CoreG (IC~4296)
(Fig. \ref{hstnuc}, right panel) 
and for the two PlawG presented in this same figure. 
The flux of the clearly detected nucleus of the CoreG IC~4296 is a factor 7
smaller than the upper limit set for the PlawG NGC~1380.

This bias in the detection threshold 
due to the different optical brightness profiles might potentially
affect the strength of the separation between the two classes 
in the diagnostic planes and in the distribution of radio-loudness
parameters.
However, we have various indications that this is not the case.

First of all, the X-ray observations are not affected by this bias
and the detection rate here is $\sim$ 90 \%. The X-ray detections also 
lead to a radio-quiet classification for UGC~5617 and UGC~7575, 
two objects lacking the optical measurement. 
The non detection of an optical nucleus in the first galaxy, 
is most likely due to nuclear obscuration since this is a 
well-studied Seyfert type 2 (AKA NGC~3226). In UGC~7575 
the optical obscuration is due to a kpc dusty disk
that is clearly visible in the HST images. 
This result emphasizes that, not unexpectedly, 
the optical data can be plagued by
the presence of absorption, accounting for some non detections.
Considering the line vs. radio luminosity diagram now, 
we have line measurements for 5 of the undetected nuclei
(indicated as stars within a circle in Fig. \ref{linepl}).
Three of them are all located in the same region of the optically detected
PlawG. This indicates that
when supporting X-ray or spectral information are available,  
they suggest in general a classification as radio-quiet AGN, even for the
objects in which the optical nucleus is undetected.

Finally, we must note that our sample can be
contaminated by the presence of non active galaxies.
In fact, we selected AGN candidates that only require
a detection above $\sim$ 1 mJy in radio observations,
but in some cases this might be produced by processes unrelated to an AGN. 
In \citetalias{paper2} we sought confirmation of a genuine AGN by
looking for the presence of i) an optical (and X-ray) core, ii) an AGN-like
optical spectrum, or iii) radio-jets.
For the CoreG we are left with 3 (out of 29) unconfirmed AGNs.
For the PlawG, 6 galaxies do not fulfill any of these criteria.
Not surprisingly, they are all at the faint end of
the radio-luminosity distribution.
Among them, for example, there is UGC~7614, the only galaxy
undetected in the X-ray, as well in the optical and in VLBI
observations, with a low radio-luminosity (Log L$_{\rm r}$=36.15) and
with only a tentative spectral classification as a transition source
between HII and LINER galaxies \citep{ho97}.

\subsection{Power-law vs. core galaxies and radio-loud vs. radio-quiet hosts}
\label{pl-cg}

The results presented above indicate that the multiwavelength
properties of the nuclei of the PlawG differ significantly from those
of CoreG, with an optical (and X-ray) excess of about 2 orders
of magnitude at equal radio-luminosity, and that a similar separation
is found comparing the emission line luminosities. 
We here explore in more detail how PlawG and CoreG
also  compare from the point of view of other parameters, such
as the host's luminosity, black-hole mass,
and optical spectra. Given the association between 
brightness profile class and radio-loudness, this 
same line of reasoning can be used to test whether
radio-loud and radio-quiet AGN differ on the basis 
of these parameters.

\begin{figure*}
\centerline{
\psfig{figure=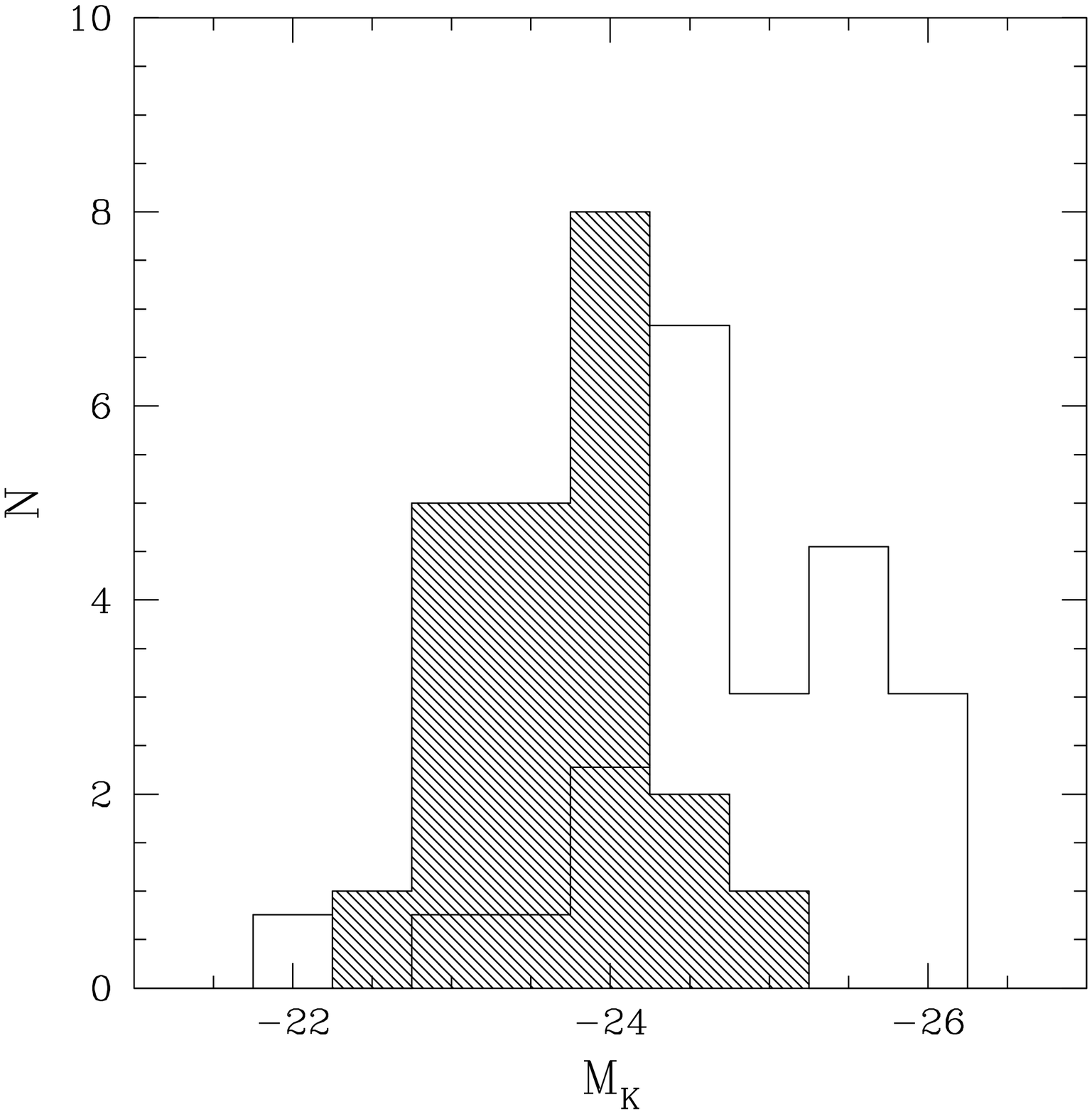,width=0.5\linewidth}
\psfig{figure=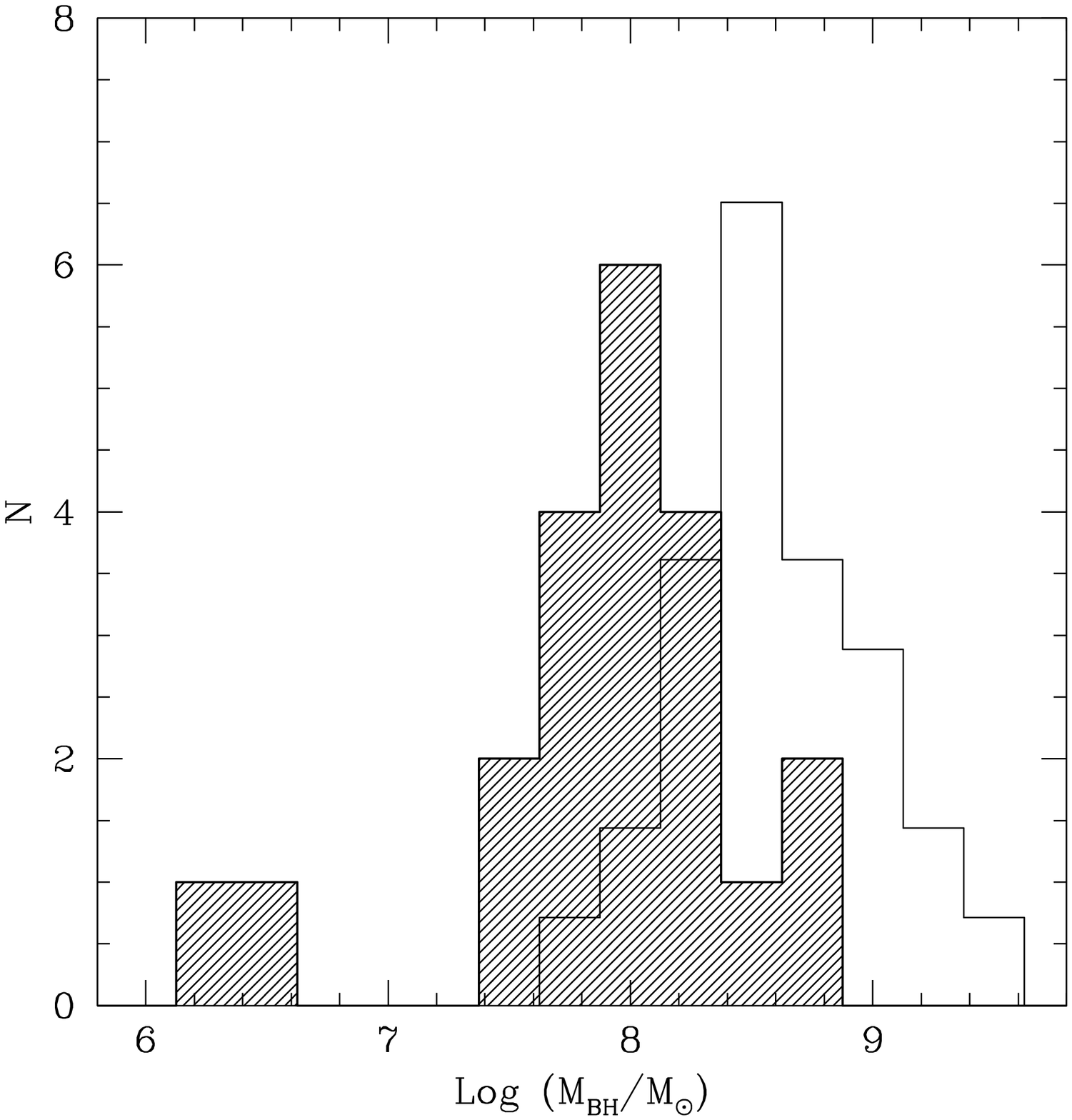,width=0.5\linewidth}
}
\caption{\label{mbhhis} Distributions for PlawG (shaded histograms) 
and for CoreG (empty histogram) of 
(left panel) absolute magnitude M$_K$ and
(right panel) black-hole mass M$_{BH}$.
The CoreG histograms have been re-normalized multiplying 
by a factor 29/22 for M$_{BH}$ and 29/21 for M$_K$, respectively, 
i.e. the number of objects
in the two samples for which estimates of these parameters are available.}
\end{figure*}

The PlawG are $\sim$ 1 magnitude fainter than CoreG,
(see Fig. \ref{mbhhis} left panel), with median values of
$M_K=-23.8$ and $M_K=-24.8$ for PlawG and CoreG, respectively. 
This reflects the well-known result that 
the brightest early-type
galaxies only show ``core'' profiles. However, 
the two classes coexist below $M_K$ = -25.

To compare their black-hole masses, when no direct measurement 
\citep[taken from the compilation by][]{marconi03} 
was available, we estimated $M_{BH}$ using
the relationship with the stellar velocity dispersion
(taken from the LEDA database) 
in the form given by \citet{tremaine02}.
The distributions of $M_{BH}$ of
the two samples are compared in Fig. \ref{mbhhis}. PlawG
have a median value, Log $M_{BH} = 8.0 $,
which is about a factor of 3 lower than for CoreG, Log $M_{BH} = 8.5 $. 
While no CoreG has a black-hole mass smaller than  Log $M_{BH} = 7.6$,
among PlawG there are two outliers from the PlawG distribution,
UGC~8675 and UGC~12759, with estimated masses of a few 10$^6$ solar masses.
As already found for the host's luminosity, however, a large overlap
between the black-hole masses is present. 

Considering the optical spectra, among PlawG we have 3 objects
classified as Seyfert and 11 as LINER. For the
CoreG, we found 13 LINER and one galaxy 
with diagnostic line  ratios that are on the borderline
between LINER and Seyfert. This, on the one hand,
confirms the result obtained by
\citet{chiaberge05}
that a dual population is associated with galaxies with a LINER
spectrum, being formed by both radio-quiet and radio-loud objects.
On the other the bulk of the two sub-samples of early-type galaxies 
are associated with similar low-ionization optical spectra,
while possibly Seyfert nuclei are hosted only by PlawG.
Therefore, no clear 
subdivision of early-type galaxies can be obtained relying 
on the host galaxies luminosity, black  hole mass, or optical spectrum.

\begin{figure*}
\centerline{
\psfig{figure=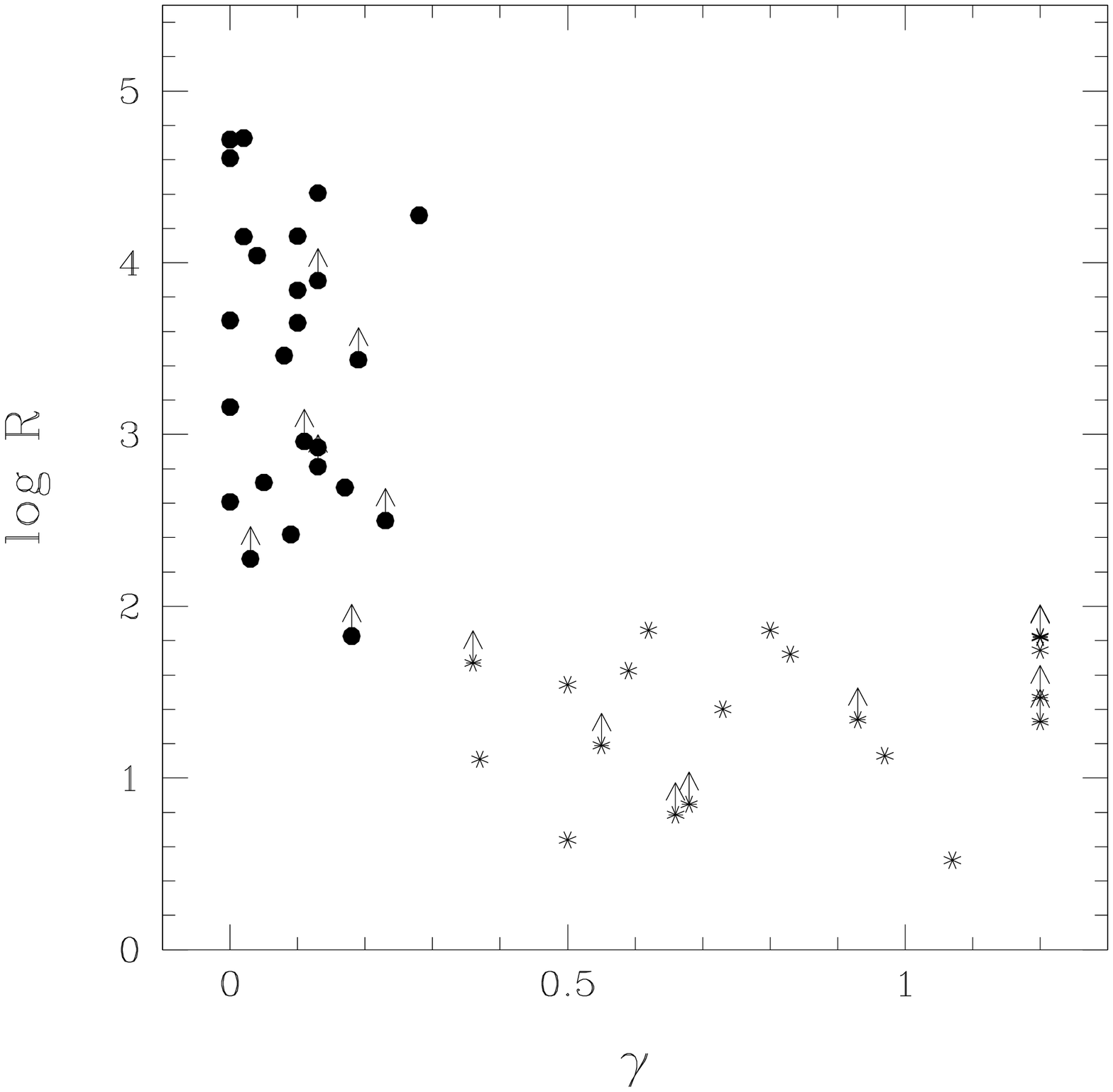,width=0.333\linewidth}
\psfig{figure=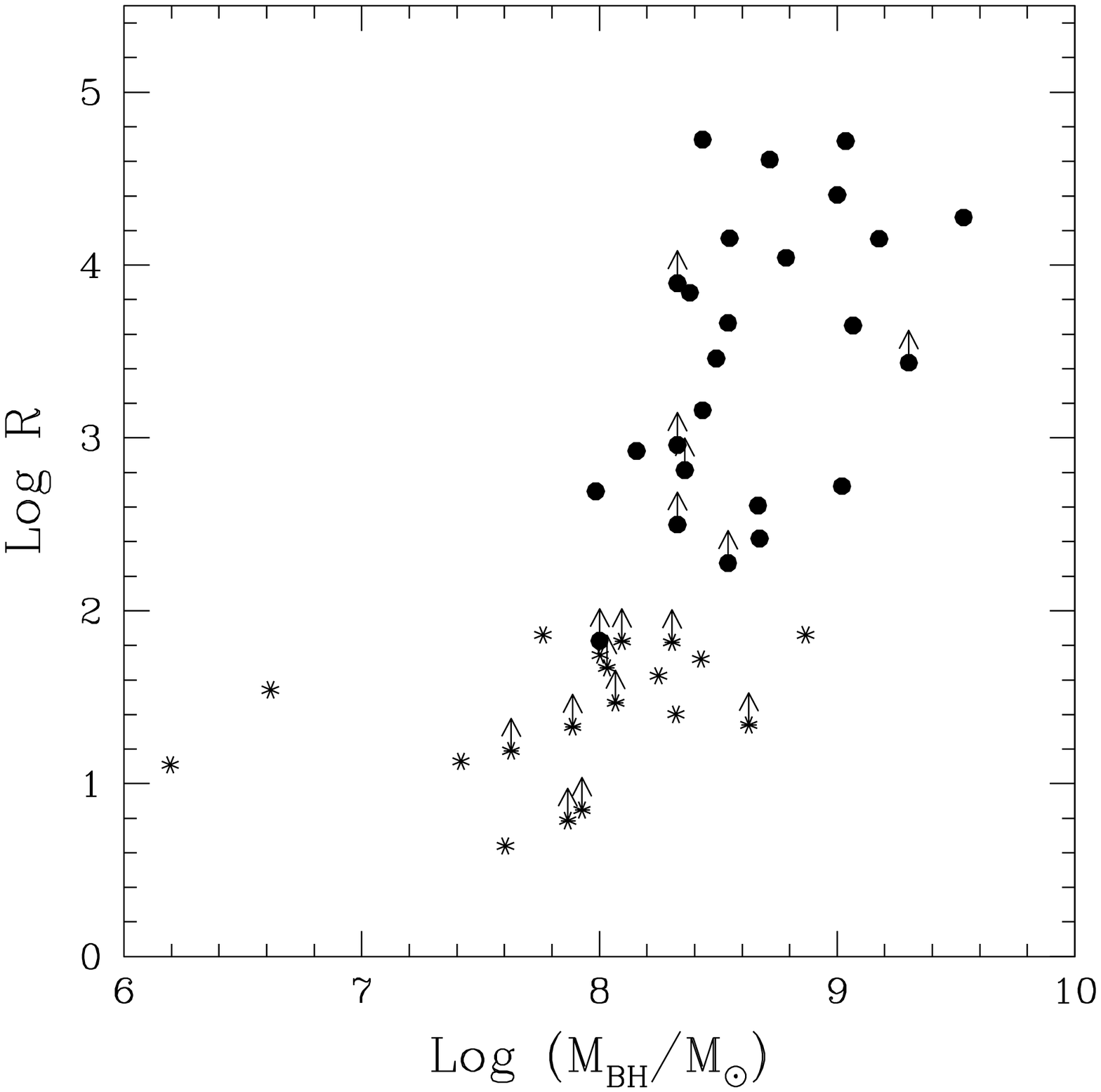,width=0.333\linewidth}
\psfig{figure=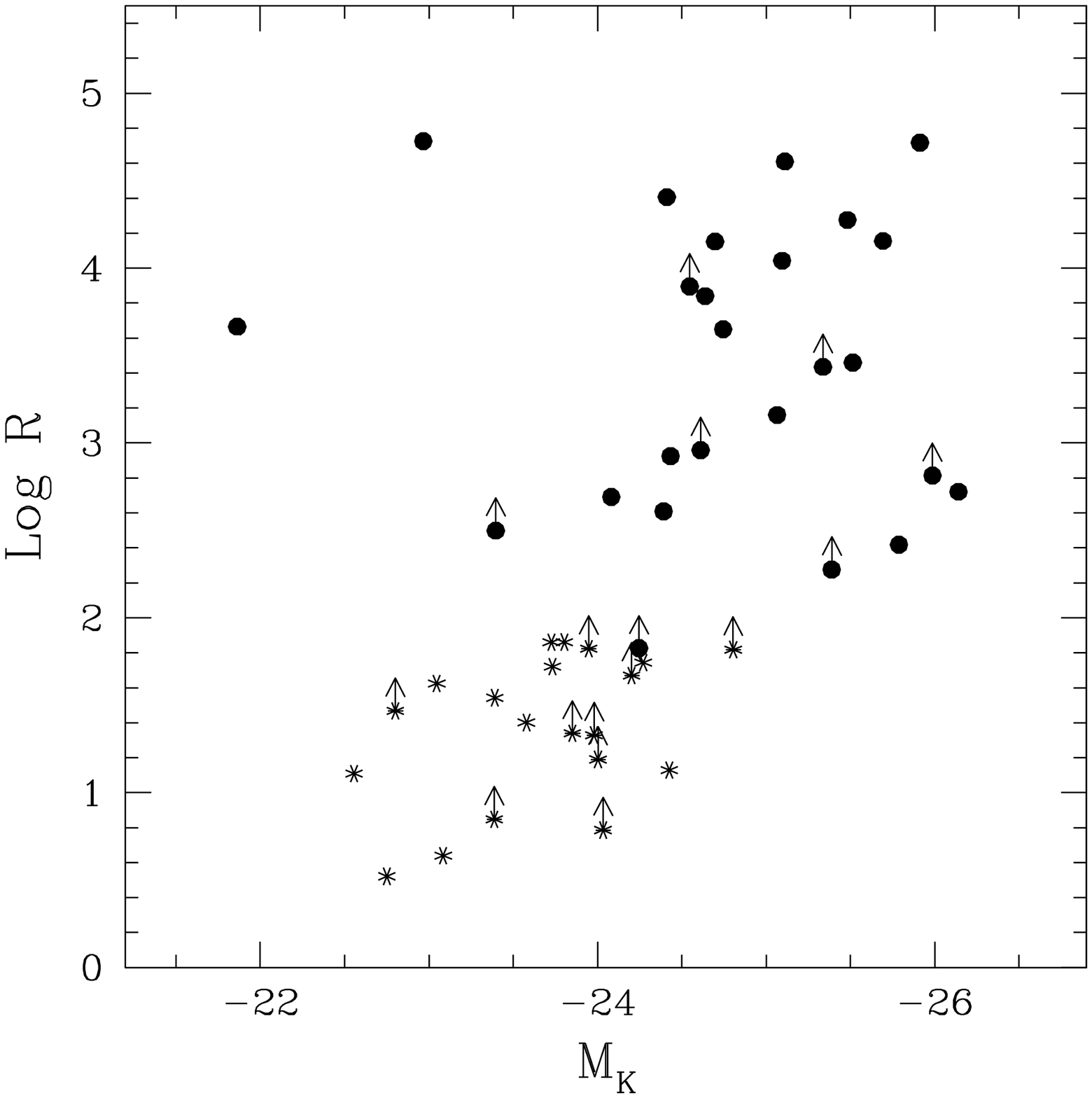,width=0.333\linewidth}}
\caption{\label{rloudom} Radio-loudness parameter R vs. (left) 
estimated black-hole mass and (centre) galaxies absolute magnitude.
Stars represent power-law galaxies, circles mark core galaxies.
No clear separation between radio-loud and radio-quiet nuclei can be
obtained based on these two quantities alone. (Right) radio-loudness
vs. $\gamma$, the logarithmic slope of the brightness profile.
Power-law galaxies in
which no break in the brightness profile is seen above the
resolution limit are set arbitrarily to $\gamma=1.2$.}
\end{figure*}

This result can be extended to a possible link of these parameters to 
the radio-loudness of AGN. 
This point is made clearer by Fig. \ref{rloudom}, in which we plot
the optical radio-loudness parameter against the black-hole mass
and the host's luminosity. Although the most massive 
black holes are associated only with radio-loud nuclei,
and the 2 least massive ones to radio-quiet AGN, 
both classes are found for the range 
8 \ltsima Log M$_{BH}$ \ltsima 9. 
A possible distinction based on the galaxy's magnitude
is even less clear, since it only emerges that
galaxies brighter than $M_K$ \ltsima -25 exclusively
host radio-loud AGN: below this level both 
classes of radio-loudness are found down to the luminosity limit 
of the sample, with the least luminous galaxy associated with
a radio-loud nucleus.

Consequently, the only
univocal connection linked to the different radio-loudness 
class of the AGN hosted by early-type galaxies 
is found when considering their brightness profiles. This is clearly
shown by the right panel in 
Fig. \ref{rloudom}, where we report radio-loudness against $\gamma$, 
the logarithmic slope of the brightness profile. This figure also
shows that, when the two classes are separated, 
there is no dependence of $R$ on the nuclear slope.

\section{Constraints on the accretion process.}
\label{adaf}

\begin{figure*}
\centerline{
\psfig{figure=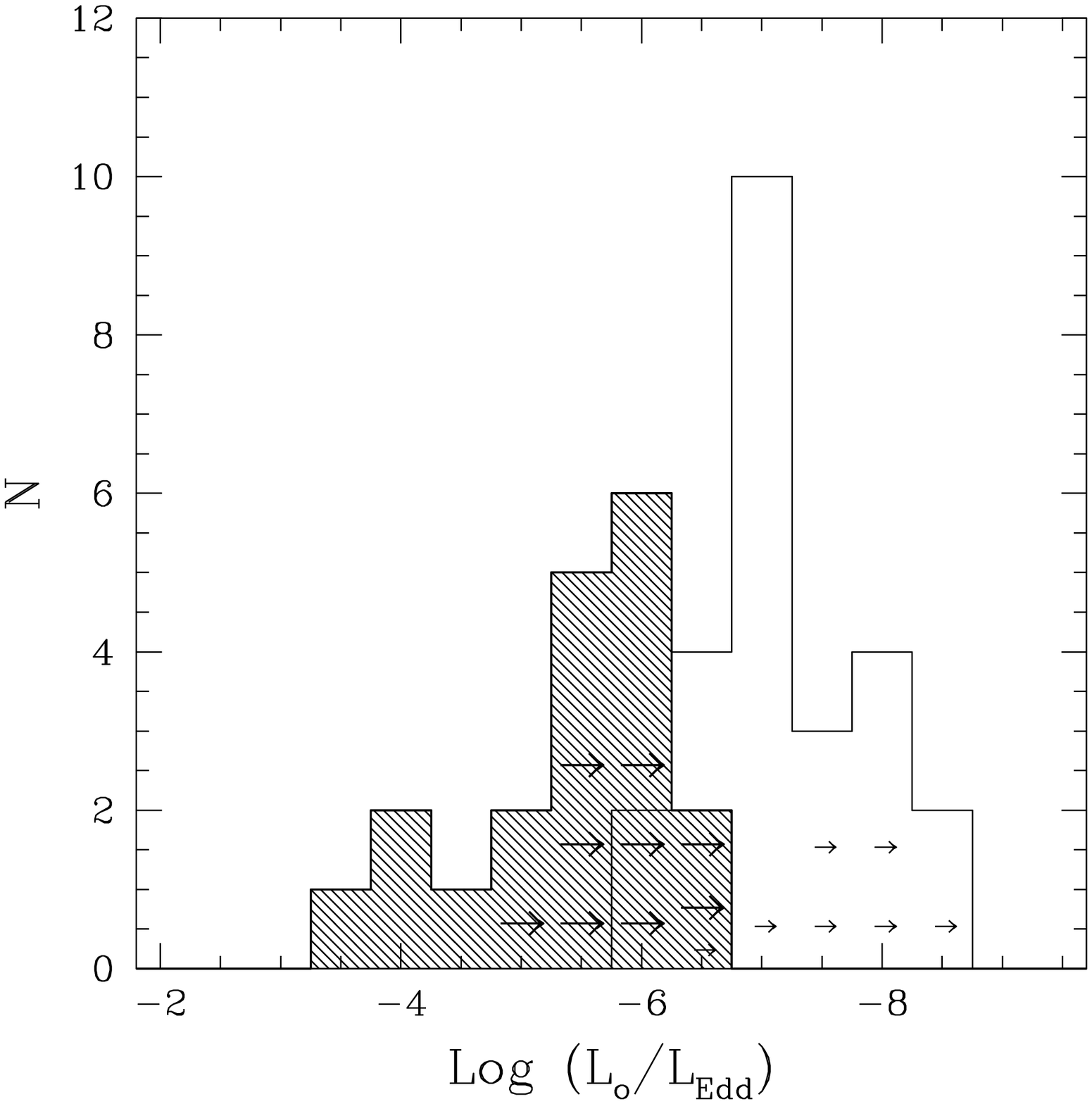,width=0.50\linewidth}
\psfig{figure=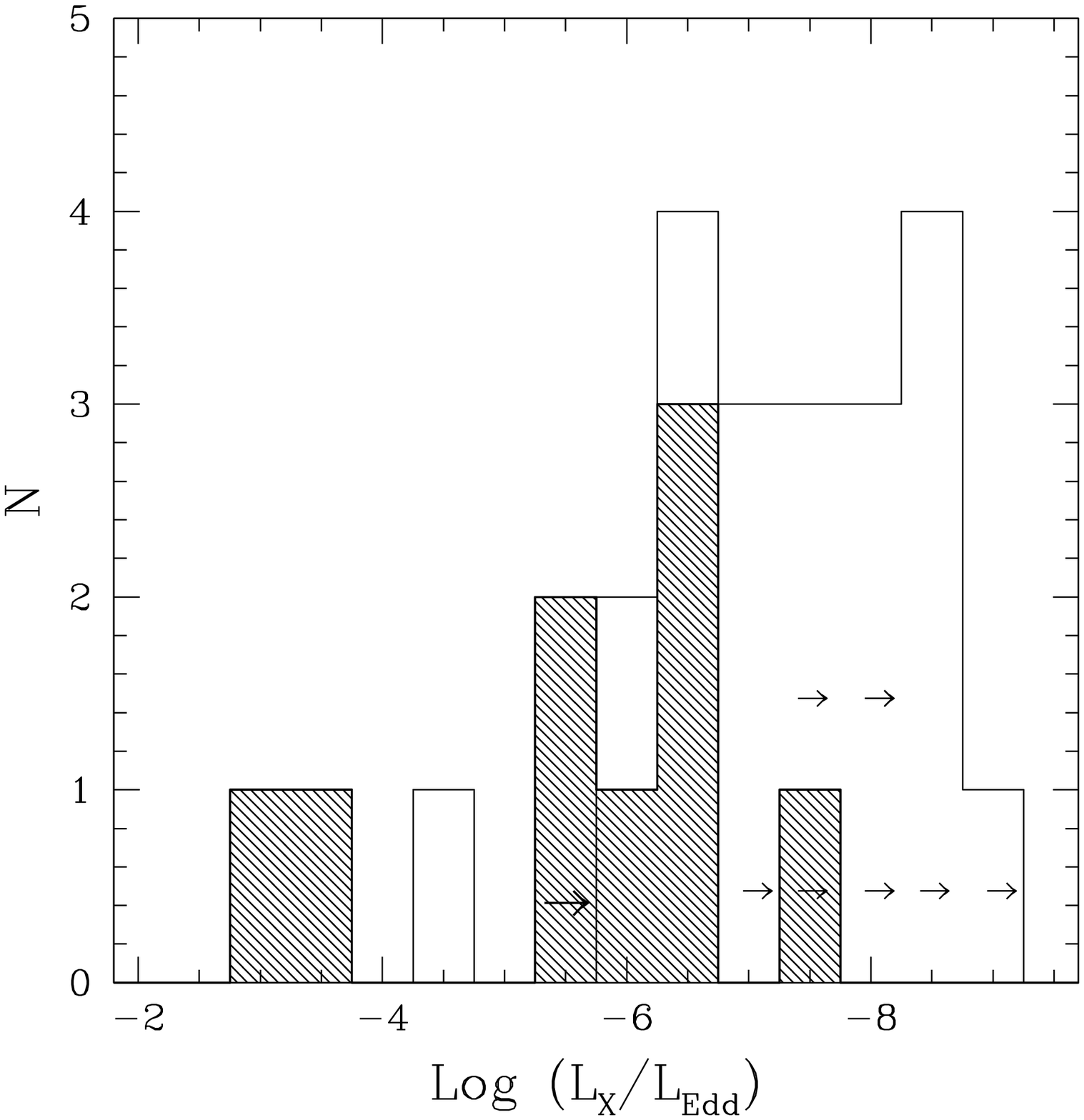,width=0.50\linewidth}
}
\caption{\label{eddihis} Distributions of the nuclear luminosities 
measured as fractions
of the Eddington luminosity in the optical (left) and the X-ray (right) 
bands for the PlawG (shaded histogram) and CoreG (empty histogram).
Upper limits for PlawG are marked with larger and heavier
arrows.}
\end{figure*}

Taking advantage of the estimates of black-hole mass we 
can refer the measurements of
the nuclear luminosities to units of the Eddington luminosity. 
PlawG nuclei are associated with a fraction of $L_{\rm {Edd}}$ 
covering a large range, $L/L_{\rm {Edd}} \sim 10^{-3} - 10^{-7}$,
in both the X-ray and optical bands 
(see Fig. \ref{eddihis}), extending down to the
regime characteristic of advection dominated accretion flows 
\citep[ADAF,][]{narayan95,narayan04}. When compared to
the CoreG, PlawG 
correspond to higher values of $L_{\rm {Edd}}$;
indeed, the two distributions match  closely
when a shift of about 2 orders of magnitude is applied.
This reflects the difference in their nuclear luminosities,
which is further enhanced by the slightly higher M$_{\rm BH}$ 
values for CoreG.

The differences between the properties of the two classes
are even more prominent considering that in CoreG the correlations between
radio, optical and X-ray luminosities strongly suggest 
that most likely we are seeing emission produced by an outflow.
For this class the measured Eddington ratios should 
probably be considered as upper limits to any emission from the accretion disk.
In PlawG the non-thermal jet radiation is likely to represent a negligible 
contribution to the light seen in the optical and X-ray. The observed nuclear
light most likely represents the genuine radiative output of the accretion
process. 

For the CoreG with the least luminous X-ray nuclei, 
the analysis performed by \citet{pellegrini05} shows 
that these can be explained
only if an important role is played by outflows (or 
by convection) to substantially suppress
the amount of gas actually reaching the central object.
In fact, their extremely low luminosity levels  
appear to contrast even with the estimates of the fiducial minimum
accretion rate 
derived for the case of spherical accretion.

The constraints for PlawG are clearly less stringent.
This is firstly due to the substantially higher
nuclear luminosities in Eddington units. Furthermore, the spherical accretion
rate scales as $\dot{M_B} \propto \rho T^{-3/2} M_{BH}^2 \propto \rho
M_{BH}^{1.44}$ (having used $T \propto \sigma^{1.5}$ 
as a proxy for
the temperature and $M_{BH} \propto \sigma^{4.02}$) where 
$\rho$ and $T$ are the density and temperature of the circumnuclear
gas. Unfortunately, detailed studies 
of X-ray data, potentially leading to estimates of $\rho$ and $T$,
are not available for PlawG. However, unless the central densities are
substantially enhanced in PlawG with respect to CoreG, the minimum
accretion rate is expected to be lower for this first class of
objects. 

Therefore, it appears that only in CoreG is it necessary to advocate
an outflow to reduce the accretion rate to a level
compatible with the low nuclear luminosities. Interestingly,
this is the same class for which the widespread detection of radio-jets
indicates the presence of a highly collimated outflow. We speculate
that a common mechanism might drive both phenomena of ejection. 

\section{Discussion and conclusions}

The aim of this series of three papers has been to explore
the classical issue of the connection between host galaxies and AGN,
in the new light shed by the recent developments 
in our understanding of the nuclear regions of
nearby galaxies. 

In \citetalias{capetti05} we selected two samples of nearby galaxies comprising
332 early-type objects. We derived a
classification into core and power-law galaxies, based
on the nuclear slope of their optical brightness profiles measured from the
available HST images. This has been possible for a sub-sample of 51 
AGN candidates, selected by imposing the detection of a radio source
at a flux limit of $\sim$ 1 mJy at 5 GHz. 

In \citetalias{paper2} we focused on the 29 ``core'' galaxies.
We used HST and Chandra archival data to isolate the nuclear
emission of these galaxies in the optical and X-ray bands, 
thus enabling us (once combined with the radio data) to study the
multiwavelength behaviour of their nuclei.
We found that core galaxies invariably host a radio-loud nucleus,
with an average radio-loudness of  
Log R = $\sim$ 3.6. 
The X-ray data provide an independent
view of their multiwavelength behaviour leading to the same result,
with an X-ray-based radio-loudness parameter of 
Log R$_{\rm X}$  = Log$(\nu L_{\rm r}/L_{\rm X})  \sim -1.3$.

In this third paper, we report
the analysis of the properties of power-law galaxies. PlawG show an excess
of about 2 orders of magnitude for their optical and X-ray
nuclear luminosity with respect to the correlation defined
by CoreG and LLRG at a given radio-core
power. This translates into substantially lower
radio-loudness parameters with median values of 
Log R $\sim$ 1.6 and Log R$_{\rm X} \sim$ -3.3.  
These values are similar to those measured in Seyfert galaxies.

Thus the radio-loudness 
of AGN hosted by early-type galaxies appears to be univocally 
related to the host's brightness profile: radio-loud AGN are
only hosted by core galaxies while radio-quiet AGN are found only
in power-law galaxies. 

Radio-loud and radio-quiet nuclei 
in our sample cannot be distinguished
on the basis of other parameters, such as the host's luminosity,
the black-hole mass (they differ only on a statistical basis),
or the optical spectral classification. 
This reflects the situation, discussed in the Introduction, 
encountered in the past when trying to understand the 
connection between host galaxies and AGN.
Only the brightness profiles provide a full separation between the
two classes.

In the Introduction we also reported 
that spiral galaxies only harbour radio-quiet
AGN \citep[with the notable exception of the radio-source
0313-192,][]{ledlow98}. 
\citet{seigar02} shows that spiral galaxies in which the large scale spheroids
can be described with a R$^{1/4}$-law (as opposed to
exponential bulges) also show steep nuclear
cusps in HST images, with $\gamma > 0.5$. 
This is indicative of a similarity between these luminous bulges
and the early-type power-law galaxies.
The R$^{1/4}$-law bulges are usually found in spiral galaxies of Hubble type T
$\lta 3$ that also represent the predominant host type for Seyfert
galaxies \citep[e.g.][]{maia03}.
This fits nicely with our finding that radio-quiet nuclei
are associated with galaxies with a power-law brightness profile
and suggests an extension of this link also to spiral hosts. 

A  possible interpretation of the connection between the brightness
profile and the AGN radio-loudness relies  on the  different
formation processes of core and  power-law galaxies. In fact, it has
been suggested  that a  core galaxy  is the result  of (at  least) one
major merger and  that the core formation is  related to the dynamical
effects of the binary black holes on the stellar component 
\citep[e.g.][]{milo02,ravi02}.  
Conversely, power-law galaxies  partly preserve  their original
disky  appearance, suggestive  of  a series  of 
minor gas-rich mergers \citep{faber97,ryden01,khochfar03}.  
The surface brightness profile
thus results from the formation history  of the galaxy, via mergers.  
The merger process also involves the SMBH associated with each galaxy
that rapidly  sink toward  the centre  of  the forming
object \citep{begelman80}. 
The connection between the properties of the galaxy 
(i.e. its mass and velocity dispersion) and
the black-hole mass sets a link between the characteristics
of the newly formed galaxy and the system of binary black holes
at its centre.
The resulting nuclear configuration after  the merger (described  by e.g.
the total  mass, spin, 
mass  ratio, or separation  of the SMBHs)  is then directly
related to the evolution of the host. These same parameters
are most likely at the origin of the different levels of
the AGN radio-loudness.

Several different aspects of the merger process
must be explored to take full advantage of our result
as to explore the origin of the radio-loud/radio-quiet dichotomy,
as well as to provide a further tool for exploring  
the co-evolution of galaxies  and supermassive black holes
in this context.
In particular, it is still unclear whether, and under which
conditions, the black holes at the centre of the newly-formed galaxy
coalesce \citep[e.g.][]{makino04}. 
This has a very important impact on the AGN physics. 
For example, among the different
viable  interpretations,  \citet{wilson95}
suggested that a radio-loud source can form only after the coalescence
of two SMBH of similar (large) mass, forming a highly spinning nuclear
object, from which the energy necessary to launch  a relativistic jet
can be extracted. Note that, 
in this situation (the merging of two large galaxies
of similar mass) the expected outcome is a massive core-galaxy,
in line with our results. 

Substantial progress in our understanding 
of the dichotomy of the brightness profiles is also required,
mainly from the point of view of the relationship between the galaxies 
formation
history and their brightness profiles. Detailed N-body simulations 
exploring a wide range of masses and mass-ratios between the merging
galaxies, including the effects of the black holes,
are still not available. But the situation
is far from settled also from the point of view 
of modeling the observed brightness profiles. For example, 
the Nuker's fitting strategy has been challenged by
\citet{graham03}. They argue that a S\'ersic model
\citep{sersic68} characterises 
the brightness profiles of early-type galaxies better.
It is unclear whether the dichotomy in early-type galaxies 
found by \citet{faber97} is preserved when this different strategy 
is adopted. When our sample is concerned, the situation appears
to be rather clear,
since all objects we classified as core-galaxies  
are recovered as such according to the Graham et al. 
definition too. This is probably due to the fact that we are dealing
with nearby galaxies, with well-resolved cores, and covering 
a rather limited range in absolute magnitude. 

From our data we cannot establish whether
the separation between radio-loud and radio-quiet
objects and the relationship with the brightness profile is
associated with a smooth transition or whether it is
the manifestation of a threshold effect. 
From the point of view 
of the radio-loudness measurements, although PlawG and CoreG
are well separated, there is no gap in their radio-loudness 
distributions. In part this is  due to the scatter (in
addition to the dispersion intrinsic of each class) 
introduced by measurement errors
and by nuclear variability, since the multiwavelength
data are not simultaneous.
The relatively small number of objects with complete 
multiwavelength coverage and the presence of upper limits 
inhibits a detailed analysis of this issue. 
The lack of a dependence of $R$ on the
nuclear slope (once PlawG and CoreG are separated) 
argues against the existence of a
continuity between the two classes of early-type galaxies.
Furthermore, at least one of our selection criteria, i.e. the detection
of a radio source, biases the sample toward the inclusion 
of the radio-quiet AGN with higher radio-loudness parameters.
In this sense our sample of low-luminosity AGN
is likely to represent only the tail, toward high values of $R$, for the 
overall population of radio-quiet AGN. If this is the case,
the separation between the two classes might be even more pronounced. 
It is interesting that even for QSOs the issue of the distribution
of radio-loudness parameters is still a matter for debate, in particular 
concerning the influence of the biases introduced by the methods of sample
selection on the observed dichotomy \citep[see e.g.][]{cirasuolo03}.
 
Our data were also used to explore 
the radiative manifestation of the accretion process
onto the supermassive black holes hosted by early-type galaxies.
CoreG have nuclear luminosities $\sim$ 2 orders of magnitude
fainter than PlawG. This is somewhat surprising since
there is evidence that in CoreG there is a 
substantial contribution from jet emission,  
which should make them brighter, everything else being equal, than
PlawG. Since the least luminous PlawG are already in the regime
characteristic of low efficiency accretion flows, e.g. ADAF, 
this difference is most likely due to a lower accretion rate in
CoreG. We have speculated 
that a common mechanism might drive both the relativistic jets
commonly observed in the radio-loud CoreG, as well as 
an outflow (originating at larger radii), which reduces the 
accretion rate to a level
compatible with their low nuclear luminosities. 

Our analysis is based on a sample of nearby
galaxies (since only in galaxies at relatively small redshift
is it possible to separate galaxies
with different brightness profiles with high confidence) and 
this results in the selection of AGN of low luminosity.
It would be very interesting to assess whether the association
between AGN radio-loudness and host profile can be
extended to more luminous AGN. 
In \citet{deruiter05} we demonstrated that 
radio-galaxies (with both FR~I and FR~II morphology), 
on average far brighter than those found in our sample, 
are hosted exclusively by core galaxies, indicating that our scenario applies 
to radio-loud AGN with a radio-luminosity as large as
 L$_r \sim 10^{42}$ erg s$^{-1}$. For the radio-quiet objects, 
the case of Seyfert galaxies is very promising. Not only is there 
a substantial population of nearby Seyfert galaxies hosted by
early-type galaxies,
but in the case of Type 2 galaxies the nucleus
is obscured at optical wavelengths. This leaves open the possibility
of exploring their nuclear brightness profiles in detail and to test the
prediction that they should be hosted by power-law galaxies.
Unfortunately this cannot be expected, at least in the near future, 
for quasars that combine larger distances
and even brighter nuclei than Seyfert galaxies. 
However, a wider applicability of the
connection between AGN properties and brightness profiles, might 
provide useful insights also for the interpretation of the 
results obtained from the study of QSO host galaxies.

The results presented here can be put on stronger ground 
with new observations. For example, the X-ray data provide us
with the clearest separation between the two classes. 
This is due to the combination of two effects with
respect to the optical data: 1) the X-ray data are not
subject to the bias of detecting faint nuclei in galaxies 
with high optical brightness and 2) they are less affected 
by nuclear obscuration.
Observing time with the Chandra telescope has been recently awarded
to us to obtain full coverage with X-ray data of the Northern sub-sample.

Similarly, the available radio data are often not adequate for isolating
and measuring the radio-cores accurately. 
For example, VLBI observations of the PlawG UGC~5663  
provided an upper limit of 0.5 mJy (Filho et~al. 2004)
to be compared with the 3.3 mJy VLA flux; 
its radio-loudness parameter, the larger among the
radio-quiet nuclei, is then actually an upper limit.  
But, most important, the lack of high sensitivity radio observations
completely inhibits the multiwavelength study 
of all galaxies undetected in the radio surveys. 
The detection of fainter radio-sources associated
with early-type galaxies has the potential to expand the sample 
with complete multiwavelength coverage by a factor of 3
and, at the same time, to extend our study to an even lower level 
of radio power.

Substantial progress can also be expected from
new HST data. First of all, 
they can lead to a Plawg/CoreG classification for a larger
fraction of objects in the sample, currently only at a $\sim$
44\% level. Near-infrared HST observations are likely to provide 
useful brightness profiles in the significant fraction of objects
whose optical structure is plagued by the presence of dust features.
Finally, observations at shorter
wavelength, where the contrast between the nuclear emission
and the galactic starlight is enhanced,  
might allow the detection of a higher fraction of optical
sources with respect to the available images.

\begin{acknowledgements}
This work was partly supported by the Italian MIUR under 
grant Cofin 2003/2003027534\_002.
This research made use of the NASA/IPAC Extragalactic Database (NED
operated by the Jet Propulsion Laboratory, California Institute of
Technology, under contract with the National Aeronautics and Space
Administration), of the NASA/ IPAC Infrared Science Archive
(operated by the Jet Propulsion Laboratory, California
Institute of Technology, under contract with the National Aeronautics
and Space Administration) and of the LEDA database.
\end{acknowledgements}

\appendix

\section{Notes on the X-ray observations of the individual sources}
\label{notes}
In this Appendix we list references and provide comments for the Chandra data
and X-ray nuclear measurements found in the literature.

\begin{figure}
\centerline{
\psfig{figure=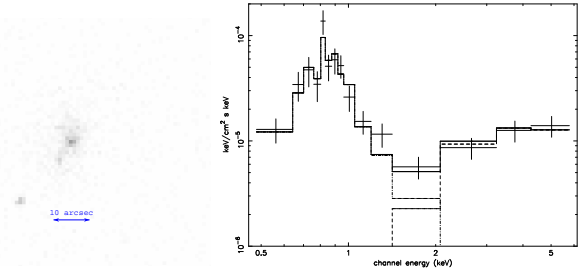,width=0.42\linewidth}
\psfig{figure=4490fa1b.ps,width=0.54\linewidth,angle=270}}
\caption{\label{fit1}
Chandra  image and  spectrum  for UGC~7103.
The fit and  the contributions of  the two
components (thermal and power-law) are also plotted.}
\end{figure}

\begin{figure}
\centerline{
\psfig{figure=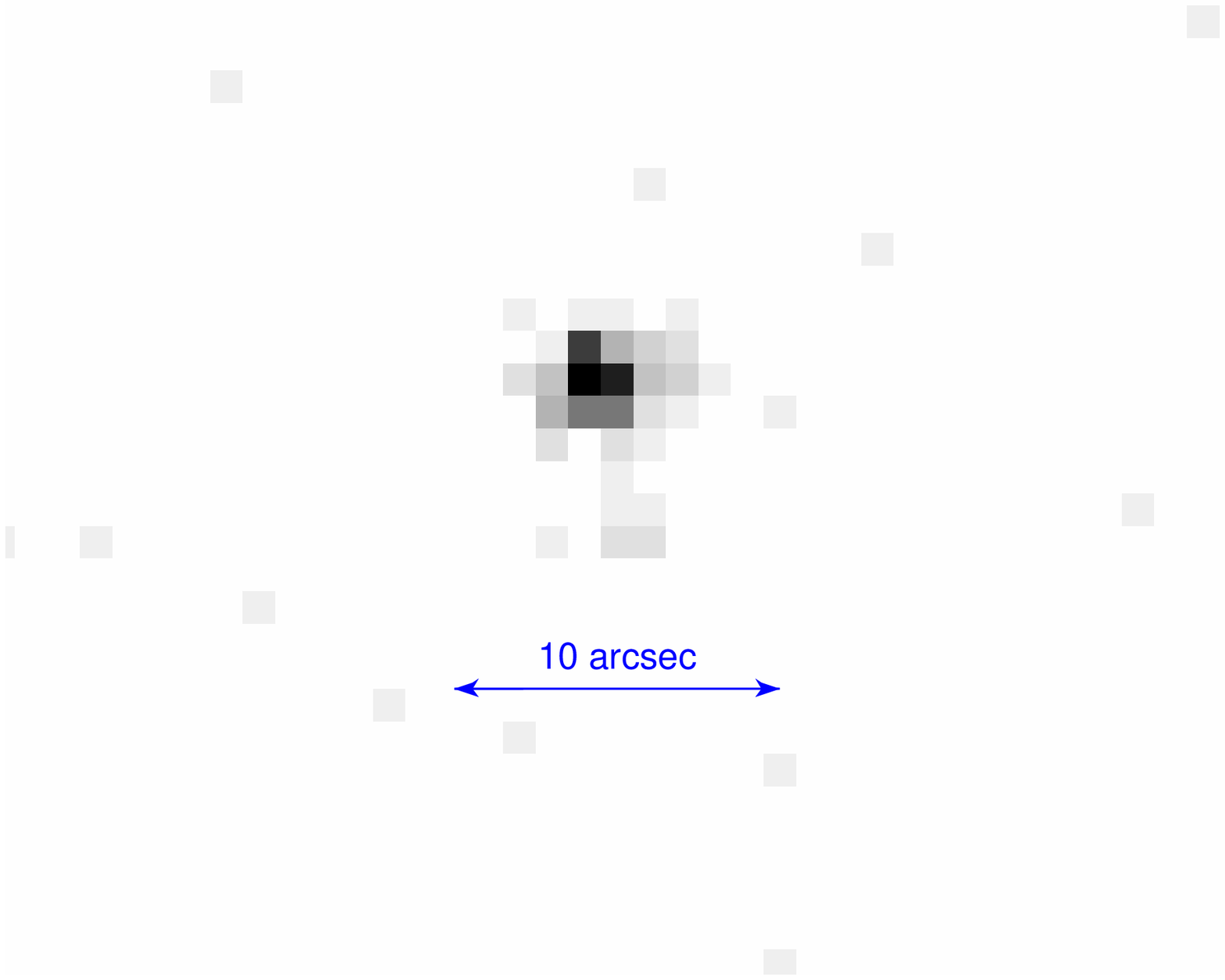,width=0.42\linewidth}
\psfig{figure=4490fa2b.ps,width=0.54\linewidth,angle=270}}
\caption{\label{fit2} 
Chandra  image and  spectrum  for UGC~8675.
The fit and  the contributions of  the two
components (thermal and power-law) are also plotted.}
\end{figure}

We also summarise of the results of the data analysis
for the unpublished Chandra observations for 3 galaxies 
in Table \ref{newchandra}.
We analysed these
observations following the same strategy as in
\citet{paper2}. Briefly, using the Chandra 
data analysis  CIAO v3.0.2,  with the
CALDB version 2.25, 
we reprocessed all the data from level 1 to level 2,
subtracting the bad pixels,  applying ACIS CTI correction, coordinates,
and pha randomization.  We searched for background flares and eventually
we excluded some period of bad aspect.
We extracted the spectrum in a circle region centred on the nucleus 
with radius of 2\arcsec, measuring the background in an annulus
of 4\arcsec radius.  We then
fit the spectrum using an absorbed power-law plus a thermal
model,  with  the  hydrogen  column  density  fixed  at  the  Galactic
value. In Figs. \ref{fit1} and \ref{fit2} 
we give images and spectra for the 2 newly detected
X-ray nuclei.
 
\begin{table*}
\caption{
\label{newchandra} 
Summary of the results of the data analysis
for the 3 unpublished Chandra observations}
\begin{tabular}{l | c c c c | c c c c c}
\cline{1-10} 
& \multicolumn{4}{|c|} {Observation information} & \multicolumn{5}{c}{Fit results} \\
Name & Obs Id & date & Inst & Exp time & N$_{H}^{z}$ & $\Gamma$ & KT & F$_{x,nuc}$(1 keV) & $\chi^2$/d.o.f or PHA bins \\
\hline
UGC~7103 & 1578   &  2001-04-03   & ACIS-S & 15.0   & 4.6$_{-1.8}^{+3.2}$E22 & 1.9 (fixed) & 0.6$_{-0.1}^{+0.1}$ & 1.0$_{-0.3}^{+0.4}E-13$ & 7.3/15 \\
UGC~7614 & 2927   &  2002-06-02   & ACIS-S & 10.0   & --                       & 1.9 (fixed) &         --          & $<1.9\;E-14$          & -- \\
UGC~8675 & 415    &  2000-09-03   & ACIS-S & 2.0    & 4.4$_{-1.1}^{+1.4}$E22 & 1.9 (fixed) & 0.4$_{-0.1}^{+0.3}$ & 8.2$_{-1.8}^{+3.3}E-13$ & 9 PHA bins   \\
\hline
\end{tabular}
\end{table*}

\noindent
{\bf UGC~5617:} 
The spectrum of the nuclear source is fitted with an absorbed power-law with a photon index $\Gamma= 1.94_{-0.25}^{+0.26}$ and
an intrinsic absorption of $N_H=4.8\; 10^{21} $cm$^{-2}$ \citet{george01}.

\noindent
{\bf UGC~5663:}
\citet{filho04} find a hard nuclear X-ray source,spatially
coincident with the optical nucleus.
The spectrum of the nuclear source is fitted 
with a power-law ($\Gamma$ fixed at value 1.7).

\noindent
{\bf UGC~6153:}  
\citet{netzer02} fitted the
spectrum obtained with 
the Chandra Low  Energy Transmission
Grating using a photon  index $\Gamma$=1.7 plus a reflection spectrum
of  gas with  a  column  density of  N$_H$=10$^{24}$  cm$^{-2}$ and  a
covering factor of 0.4-0.7.

\noindent
{\bf UGC~7103}:   We  fitted   the  spectrum   with  an   absorbed  powerlaw
(N$_H =4.6\;10^{22}$ cm$^{-2}$  
and a photon index  $\Gamma$ fixed to 1.9) plus a thermal model.

\noindent
{\bf UGC~7142:}
In a circle of radius 2$\arcsec$ \citet{terashima03} obtained total
nuclear counts of 32.6 in the 2-8 keV range.
A power-law model model modified by absorption was applied to the data
obtaining $\Gamma=1.66$ and $N_H=1.5\;10^{20} cm^{-2}$. 

\noindent
{\bf UGC~7256:}  
\citet{ho01} measured 410 nuclear counts from 
a 2 \arcsec diameter circle. 
The ACIS count rate are converted to x-ray luminosity
by assuming $\Gamma$ =1.8 and $N_H=2\;10^{20} cm^{-2}$.

\noindent
{\bf UGC~7575:}
\citet{machacek04} analysed the total emission from the galaxy (within
32  \arcsec of the  nucleus).  The  data were  described by  a thermal
MEKAL plasma  plus a power-law component. Fixing  the absorbing column
at the Galactic  value (2.64$\; 10^{22} $cm$^{-2}$) and  the MEKAL model
abundance at 0.1  Z$_{\odot}$, the data are well  fitted with a photon
index $\Gamma$ =1.22.

\noindent
{\bf UGC~7614}:We  estimated  an upper  limit  to  any non-thermic  emission
fixing the photon index to $\Gamma$=1.9.

\noindent
{\bf UGC~8675}: We grouped  the data to have at least  10 counts per channel
and we applied the Poissonian statistic.  The spectrum is fitted using
an absorbed powerlaw (N$_H=4.4\; 10^{22}$ cm$^{-2}$ and photon index
$\Gamma$ fixed to 1.9) plus a thermal model.

\end{document}